\documentclass[
  journal=largetwo,
  manuscript=article-type,
  year=2020,
  volume=37
]{cup-journal}

\usepackage{aas-macros}
\usepackage[utf8]{inputenc}
\usepackage{csquotes}
\usepackage[english]{babel}

\usepackage{amsmath,amssymb}
\usepackage[nopatch]{microtype}
\usepackage{booktabs}
\usepackage{siunitx}
\usepackage{graphicx}
\usepackage{rotating}
\usepackage{multirow}
\usepackage{tablefootnote}
\usepackage{tikz}
\usepackage{upgreek}
\usepackage{overpic}

\usepackage{biblatex}
\newcommand{\DM}{\ensuremath{{\rm DM}}}
\newcommand{\DMeg}{\ensuremath{{\rm DM}_{\rm EG}}}

\newcommand{\DMmw}{\ensuremath{{\rm DM}_{\rm MWISM}}}
\newcommand{\DMhalo}{\ensuremath{{\rm DM}_{\rm halo}}}
\newcommand{\DMhost}{\ensuremath{{\rm DM}_{\rm host}}}
\newcommand{\DMcosmic}{\ensuremath{{\rm DM}_{\rm cosmic}}}
\newcommand{\muhost}{\ensuremath{\mu_{\rm host}}}
\newcommand{\sighost}{\ensuremath{\sigma_{\rm host}}}
\newcommand{\emax}{\ensuremath{E_{\rm max}}}
\newcommand{\emin}{\ensuremath{E_{\rm min}}}
\newcommand{\lemax}{\ensuremath{\log_{10} \emax ({\rm erg})}}
\newcommand{\lemin}{\ensuremath{\log_{10} \emin ({\rm erg})}}
\newcommand{\zdm}{{\sc zDM}}
\newcommand{\pth}{{\sc PATH}}
\newcommand{\pccc}{\si{pc\,\centi\metre^{-3}}}

\newcommand{\nsfr}{\ensuremath{n_{\rm sfr}}} 
\newcommand{\snr}{\ensuremath{{\rm S/N}}} 
\newcommand{\fdz}{\ensuremath{f_d(z)}}

\newcommand{\xrad}{\ensuremath{\mathbf{x}_{\rm rad}}}
\newcommand{\xopt}{\ensuremath{\mathbf{x}_{\rm opt}}}

\newcommand{\weff}{\ensuremath{w_{\rm eff}}}

\newcommand{\frb}{FRB\,20190611B}
\newcommand{\Andersen}{B.C.\,Andersen et al.\ (in prep., 2026)}
\newcommand{\marnoch}{L.\,Marnoch et al., in prep 2026}
\newcommand{\wangjames}{Z.\ Wang et al., 2016 (in prep)}
\newcommand{\hostA}{Galaxy A} 
\newcommand{\hostB}{Galaxy B}

\title{Through a glass, darkly: a combined framework for estimating fast radio burst host galaxy and population properties in an era of uncertain host identification} 

\author{C.~W.~James}
\affiliation{International Centre for Radio Astronomy Research, Curtin University, Bentley, 6102, WA, Australia}
\email[Clancy W.\ James]{clancy.james@curtin.edu.au}

\author{B.~C.~Andersen}
\affiliation{Department of Astronomy and Astrophysics, University of California, Santa Cruz, CA 95064, USA}

\author{L.~Marnoch}
\affiliation{School of Mathematical and Physical Sciences, Macquarie University, NSW 2109, Australia}
\alsoaffiliation{Astrophysics and Space Technologies Research Centre, Macquarie University, Sydney, NSW 2109, Australia}
\alsoaffiliation{Australia Telescope National Facility, CSIRO Space \& Astronomy, Box 76 Epping, NSW 1710, Australia}

\author{J.L.~Hoffmann}
\affiliation{International Centre for Radio Astronomy Research, Curtin University, Bentley, WA 6102, Australia}

\author{N.~Loudas}
\affiliation{Department of Astrophysical Sciences, Princeton University, 4 Ivy Lane, Princeton, NJ 08544, USA}

\author{J.~X.~Prochaska}
\affiliation{Department of Astronomy and Astrophysics, University of California, Santa Cruz, CA 95064, USA}
\alsoaffiliation{Kavli Institute for the Physics and Mathematics of the Universe, 5-1-5 Kashiwanoha, Kashiwa 277-8583, Japan}
\alsoaffiliation{Division of Science, National Astronomical Observatory of Japan, 2-21-1 Osawa, Mitaka, Tokyo 181-8588, Japan}

\author{S.~D.~Ryder} 
\affiliation{School of Mathematical and Physical Sciences, Macquarie University, NSW 2109, Australia}
\alsoaffiliation{Astrophysics and Space Technologies Research Centre, Macquarie University, Sydney, NSW 2109, Australia}

\author{M.~Woodland}
\affiliation{Department of Astronomy and Astrophysics, University of California, Santa Cruz, CA 95064, USA}


\begin{document}

\begin{abstract}

The identification of fast radio burst (FRB) host galaxies, and subsequently their redshifts ($z$), has allowed the FRB dispersion measure (DM) to be used to probe the cosmological distribution of ionised gas, and study the properties of the FRB population itself. However, current methods cannot account for FRBs with uncertain host galaxy associations, leading to underutilisation of data, and potential biases towards nearby, bright hosts.
In this work, we develop a methodology which can. We do so by combining three ingredients --- the \zdm\ code, for modelling the Macquart relation; \pth, for statistical host galaxy identification; and a set of models able to describe the intrinsic FRB host galaxy distribution --- into a single formalism. We prove the fidelity of our formalism by using a synthetic set of FRB observations, with simulated hosts sampled from galaxy catalogues, and show that it reproduces intrinsic host galaxy properties even for FRB localisation uncertainties of $30"$, where a traditional \pth\ analysis produces no confidant host associations. When applied to a sample of FRBs localised by the Australian Square Kilometre Array Pathfinder, we confirm previous results showing that FRBs prefer host galaxies fainter than that predicted by star-formation, consistent with an exponential surface density scaling as $0.41^{+0.13}_{-0.09}$ times the half-light radius. We encourage the application of this formalism to data-sets from other FRB-hunting instruments.
\end{abstract}

\section{Introduction}
\label{sec:intro}

The identification of fast radio burst (FRB) host galaxies is critical to both using FRBs as probes of the baryonic matter distribution of the Universe, and identifying their progenitors. Spectroscopy performed on hosts has yielded the Macquart relation, being an empirical linear correlation between FRB dispersion measure (DM) and host redshift (z) \citep{Macquart2020}, showing that the local Universe is consistent with early Universe constraints on $\Omega_b h^2$ \citep{PlanckCosmology2018}. Numerous other studies have analysed the relationship between FRB dispersion measure and host galaxy redshift, yielding constraints on the fraction of baryons in the circumgalactic medium \citep{2024ApJ...965...57B, FLIMFLAMdr1,2025NatAs...9.1226C}, the Hubble constant \citep{WuHubble22, HagstotzHubble22,james_measurement_2023}, and FRB population parameters \citep{2020MNRAS.494..665L,2021A&A...651A..63G,2023ApJ...944..105S,2025PASA...42...17H}. Simultaneously, the properties of FRB host galaxies yield information on their progenitors, with their masses, color, star-formation rates and histories, metallicities, and position of the FRB within the galaxy being analysed to determine their formation pathways, with key questions being if FRBs come from one or more populations and/or are analogous to populations of known transients \citep{Bhandari+22,gordon_demographics_2023,DSA_Sharma_sfr}, and if they correlate with time-frequency properties of the FRB itself \citep{CordesTauRedshift2022,2025PASA...42..157G}. Both directions of investigation are extremely active and are rapidly evolving --- recent reviews are more complete than the brief summary given here \citep{2023RvMP...95c5005Z,2024Ap&SS.369...59L,2026enap....5..448G}, but may already be outdated.

Key to both directions of study is the correct identification of the FRB host galaxy. To this end, the Probabilistic Association of Transients to their Hosts \citep[PATH;][]{PATH} framework was developed, being a Bayesian approach to determining the probability of a galaxy observed in optical follow-up images being the true host, given the galaxy's properties, and the FRB localisation.

In a recent work, we showed how to update the PATH framework to account for expectations based on FRB DM \citep{zdm_path_hosts}. By using an expectation for FRB redshift based on an FRB's DM, $P(z|{\rm DM})$, and a model for FRB host galaxies giving their magnitude dependence on redshift, $P(m|z)$, one can calculate both informed prior expectations for the FRB host magnitude distribution, and posteriors for each observed host candidate. This methodology allows models for the FRB host galaxy distribution to be tested against data, while accounting for incompleteness of galaxies observed in optical images. Such an approach becomes increasingly important as FRB-hunting telescopes detect more FRBs, increasing the reliance on less-sensitive wide-field optical surveys for host galaxy identification.

However, there is one very obvious flaw in our approach. Models of the FRB population use ``confident'' FRB hosts to calibrate the FRB $z$--DM relation, typically those with posterior PATH probabilities of 90--95\% or greater. With over $100$ confident hosts now identified \citep{Shannon_ICS,PastorMarazuela2025,DSA_Sharma_sfr,2025ApJS..280....6C}, it is reasonable to conclude that several identifications are incorrect. And as discussed by \citet{james_measurement_2023}, a single mis-identified FRB host lying below the Macquart relation (i.e., having a high redshift for its DM) can severely skew population analyses, since models have few ways to account for underdensities of ionised baryons (this is known as the `DM cliff' effect). Furthermore, \Andersen\ has shown that placing cuts on posterior PATH probabilities biases the posterior FRB host distribution, since it is much easier to confidently associate FRBs with bright host galaxy candidates, than the much more common --- and difficult to identify --- dwarf galaxies.

Producing a complete sample of confident FRB hosts also requires only using hosts from low-DM, well-localised FRB where identification of the host is guaranteed \citep{2025PASA...42...17H}. However, the diversity of radio localisation accuracy of FRBs spans a vast range, from $\sim 10$\,arcmin for the CHIME/FRB Catalogues \citep{CHIMECat2}, through to a few mas for the VLBI localisations of repeaters \citep{2017ApJ...834L...8M}. Optical follow-up depth spans an equally wide range, from relatively shallow, wide-field catalogues at a depth of $m_R<19.5$ \citep{2018ApJ...867L..10M,WiseSuperCosmos}, to deep JWST observations down to $29^{\rm th}$ magnitude \citep{2025arXiv250801648C}. This diversity makes constructing a complete sample of firm FRBs hosts increasing complex, and requires discarding an increasingly large fraction of data.

In order to produce a fully self-consistent model of FRB population parameters, the inherent uncertainty in FRB host galaxy identification (which depends on the intrinsic host galaxy distribution) must be incorporated into FRB population modelling. Since this in turn informs host galaxy probabilities, a methodology to simultaneously fit both FRB host galaxy and population parameters is required. The development of such a statistical methodology is the subject of this paper.

In this work, we combine the likelihoods of the observed radio properties of an FRB, encapsulated by the parameters \xrad, with the observed optical properties of a follow-up image, \xopt. In \S~\ref{sec:xrad}, we review how radio properties are used to produce an estimate for the $z$--DM distribution of FRBs using the \zdm\ code, before showing in \S~\ref{sec:xopt} how to simultaneously incorporate optical properties into a combined statistic using \pth. Useful derived values are calculated in \S~\ref{sec:derived}, and we demonstrate our method on FRB FRB~20190611B in \S~\ref{sec:190611}.
In \S~\ref{sec:sandbox}, we use simulated FRBs inserted into optical catalogues to test our analysis, before in \S~\ref{sec:craft} applying our analysis to the catalogue of FRBs and optical images detected by the Australian Square Kilometre Array Pathfinder \citep[ASKAP;][]{hotan_australian_2021}. We discuss the uses and flaws in our methodology in \S~\ref{sec:discussion}

\section{Calculation of $P(\xrad)$ and $P(z,|\xrad)$ }
\label{sec:xrad}

We first aim to calculate the probability distribution of radio observables $P(\xrad)$, and a prediction for the redshift distribution of the host, $P(z|\xrad)$. We do so in the context of the \zdm\ code, which has been progressively developed over a series of works \citep{james_zdm_2022,james_measurement_2023,2024ApJ...965...57B,JamesRepeating2023,2025PASA...42...17H,HoffmannHalo}; here, we re-iterate the procedure for clarity. However, our underlying statistical principles apply to any methodology allowing redshift predictions for FRBs based on their DMs, e.g.\ \citet{2023ApJ...944..105S}. Those readers familiar with the \zdm\ code, or similar works, could skip to \S\,\ref{sec:xopt}.

\subsection{Detection probability}

We take the primary radio properties of an FRB to be its dispersion measure, DM, signal-to-noise ratio, \snr, and radio localisation region, $\theta$; all are routinely published as part of FRB catalogues. Secondary properties, which we have incorporated into modelling within the \zdm\ code, include FRB intrinsic width, $w$, scattering time, $\tau$, beam-value at which it was detected, $B$, and number of repetitions, $N_{\rm rep}$ (if any). For simplicity, we ignore FRB repetition in this work. For reasons outlined in \citet{2019MNRAS.483.1342J}, we use 
\begin{eqnarray}
    s & \equiv & \frac{\snr}{\snr_{\rm th}}, \label{eq:s}
\end{eqnarray}
rather than \snr\ directly, as the observable property of the FRB's signal-to-noise ratio. 

The FRB effective width, \weff, is calculated as per \citet{Cordes_McLaughlin_2003},
\begin{eqnarray}
    \weff & = & \sqrt{w_i^2 + \tau^2 + \delta t^2 + w_{\rm DM}^2}, \label{eq:weff}
\end{eqnarray}
where $\tau$ is the characteristic scattering time, $\delta t$ is the search time resolution, and $w_{\rm DM}$ is the dispersion measure smearing within each frequency channel of width $\delta \nu$,
\begin{eqnarray} 
    w_{\rm DM} & = & 8.3\,{\rm ms}\,  \frac{\rm DM}{\pccc} \frac{\delta \nu}{1\,{\rm GHz}} \left(\frac{\nu}{1\,{\rm GHz}} \right)^{-3}.
\end{eqnarray}
It is possible to estimate intrinsic distributions of $\tau$ and $w_i$ using the \zdm\ code, and modelling the redshift dependence of these parameters can affect the z-dependence of the rate by $\mathcal{O}\sim10$\% \citep{2026PASA...43...38J}. However, doing so significantly reduces computational speed --- and \weff\ can be estimated more robustly than the individual components $w_i$ and $\tau$. Hence, in this work, we consider only  $s,B,\weff,\DM \in \xrad$. These properties allow the intrinsic fluence of an FRB, $F$, to be calculated,
\begin{eqnarray}
  F  & = & s \frac{F_{\rm 1ms}}{B} \left(\frac{\weff}{\rm 1\,ms}\right)^{0.5}, \label{eq:F}
\end{eqnarray}
where $F_{\rm 1ms}$ is the characteristic fluence threshold of the experiment. This is defined as the fluence required for an FRB to produce a threshold \snr\ value of $\snr_{\rm th}$ at beam centre ($B=1$) --- this value is commonly quoted by FRB detection experiments at a normalised time-duration of 1\,ms.

At a given redshift $z$, this fluence implies an isotropic-equivalent energy of
\begin{eqnarray}
    E & = & \frac{4 \pi L_d^2}{(1+z)^{2+\alpha}} \Delta \nu F, \label{eq:Eiso}
\end{eqnarray}
where $\Delta \nu$ is the detection bandwidth, $L_d$ the luminosity distance, and $\alpha$ is the assumed spectral dependence of FRB fluence --- the factor $(1+z)^\alpha$ is used to normalise $E$ at the observer frequency $\nu$ in the rest-frame of the host. The probability of an FRB produced at redshift $z$ being detected in the range $s$ to $s+ds$ is therefore
\begin{eqnarray}
    P(s|z,B,\weff)  & = & \frac{L(E) \frac{dE}{ds}}{\int_{E_{\rm min}}^{\infty} L(E)dE} \label{eq:ps}
\end{eqnarray}
where $L(E)$ is the FRB energy distribution, and from Eq.~\ref{eq:F} and \ref{eq:Eiso},
\begin{eqnarray}
\frac{dE}{ds} & = & \frac{4 \pi L_d^2}{(1+z)^{2+\alpha}} \Delta \nu \frac{F_0}{B} \left(\frac{\weff}{\rm 1\,ms}\right)^{0.5}. \label{eq:ld}
\end{eqnarray}
The exponent of $2+\alpha$ is due to three factors: a factor of $\alpha$, which scales the emission energy to the normalisation frequency (here, 1.3\,GHz is used); a factor of unity, which undoes the factor of $1+z$ due to time-dilation in luminosity, which is not present in fluence, since all relevant emission times are being integrated over; and another factor of unity, which accounts for the increased bandwidth in the emission frame. Energies in \zdm\ are calculated at a fixed emission bandwidth of 1\,GHz; other works model the observable energy, and thus do not include the last factor, with the total exponent of $1+z$ in the numerator of Eq.~\ref{eq:ld} being $1+\alpha$. Our treatment is more appropriate for emission bandwidths broader than the observation bandwidth, whereas the latter treatment is more appropriate for emission bandwidths narrower than the observation bandwidth.

In this work, we use a Schechter function for $L(E)$, defined by
\begin{eqnarray}
L(E) & = & E^\gamma \exp(-E/E_{\rm cut}), \label{eq:schechter}
\end{eqnarray}
for some characteristic cut-off energy, $E_{\rm cut}$. The model also considers a minimum energy, $E_{\rm min}$, below which FRBs do not emit. All energies are normalised to $1.3$\,GHz; when the spectral-dependence $\alpha \ne 0$, Eq.~\ref{eq:ld} renormalises to this frequency. We note that the choice of an exponential for the high-energy cutoff is arbitrary; the only work deriving a non-parametric luminosity function, \citet{2025PASA...42....3A}, finds equal consistency with a pure power-law, while detailed studies of the luminosity function of strong repeaters find complex behaviour \citep[e.g.\ ][]{2021Natur.598..267L,2026MNRAS.545f1937O}.

\subsection{Volumetric rate in z--DM space}

The above section calculates the probability of observing an FRB with properties $\xrad$ given that one occurs at a specific point in z--DM space. We now calculate the rate of such occurrences. The redshift-dependence of the FRB rate is a product of the volume element
\begin{eqnarray}
V(z) & \equiv & \frac{dV}{dz \,d\Omega} \nonumber \\
 & = & D_H \frac{(1+z)^2 D_A^2(z)}{E(z)}, \label{eq:comoving_volume}
\end{eqnarray}
a time-dilation factor $d\tau/dt = (1+z)^{-1}$, and the FRB population density $\Phi(z)$ (typically expressed in units of FRBs per comoving Mpc$^3$ per proper year). This last is expressed as being proportional to the star-formation rate, modulated by a power \nsfr, i.e.
\begin{eqnarray}
\Phi(z) & = & \left( \frac{{\rm SFR}(z)}{{\rm SFR}(z=0)} \right)^{\nsfr} \label{eq:nsfr} \\
{\rm SFR}(z) & = & 1.0025738 \frac{(1+z)^{2.7}}{1 + \left(\frac{1+z}{2.9}\right)^{5.6}}. \nonumber
\end{eqnarray}
In the case that FRBs are treated as intrinsically narrow-band objects, $\alpha$ is interpreted as governing the frequency dependent rate; it is set to $0$ for purposes of calculating the intrinsic energy via Eq.~\ref{eq:ld}, but the volumetric rate $\Phi(z)$ is adjusted as
\begin{eqnarray}
\Phi(\alpha,z) & = & \Phi(z)  \left( (1+z) \frac{ \nu}{\nu_0} \right)^\alpha. \label{eq:rate_scaling}
\end{eqnarray}

The next ingredient is the DM budget, i.e.\ $p(\DM|z)$, which we decompose as per \citet{Macquart2020},
\begin{eqnarray}
    \DM & = & \DMmw + \DMhalo + \DMcosmic + \DMhost,\quad
\end{eqnarray}
representing DM contributions from the Milky Way's interstellar medium (ISM) and halo, cosmological contributions from the intergalactic medium (IGM) and the circumgalactic media (CGM) of intervening halos, and a host contribution covering the halo and ISM of the host galaxy, together with any local contribution from the FRB progenitor and surrounding environment. We take \DMmw\ from \citet{2026arXiv260211838O}, aka NE2025, and treat \DMhalo\ as a constant but variable fit parameter as per \citet{HoffmannHalo}. The expectation value for the cosmological contribution is given as per \citet{Ioka2003,Inoue2004,Deng2014},
\begin{eqnarray}
\left< \DMcosmic \right>(z)  & = & \int\limits_0^z \frac{c \bar{n}_e(z^\prime) \, dz^\prime}{H_0 (1+z^\prime)^2 {\rm E}(z^\prime)}. \label{eq:dmcosmic}
\end{eqnarray}
The mean electron density, $\bar n_e$, is calculated as
\begin{eqnarray}
\bar{n}_e = \fdz\,\rho_{b}(z)m_{p}^
{-1} \chi_{e} \label{eq:nebar}
\end{eqnarray}
where \fdz\ is the fraction of baryons in a diffuse ionised state ($\sim 85\%$), $\rho_b$ the mass density of baryons
\begin{eqnarray}
\rho_{b}(z)= \frac{3 \Omega_{b} H_0^2}{8 \pi G} (1 + z)^3, \label{eq:rhob} 
\end{eqnarray}
and $\chi_{e}= Y_{\rm{H}}+Y_{\rm{He}}/2$ the electron/baryon ratio, calculated from the primordial hydrogen and helium mass fractions $Y_{\rm{H}}$ and $Y_{\rm{He}}$. Other constants are the proton mass $m_p$, gravitational constant $G$, and baryonic density $\Omega_b$.

Eq.~\ref{eq:dmcosmic} predicts an almost 1-1 correlation between the mean value of \DMcosmic\ and $z$, now known as the Macquart relation \citep{Macquart2020}. It is almost completely degenerate in the product $H_0 \Omega_b f_d$; here, we fix the product $\Omega_b (H_0/100)^2 = 0.02242$ \citep{PlanckCosmology2018}, vary $H_0$, and take \fdz\ from the {\sc FRB} codebase \citep{prochaska_frbsfrb_2023}. We adopt a simplified cosmology accounting only for matter density $\Omega_m$ and dark energy $\Omega_\Lambda$ appropriate to the low-redshift Universe, giving
\begin{eqnarray}
{\rm E}(z) & = & \sqrt{\Omega_m(1+z)^3 + \Omega_\Lambda}. \label{eq:Ez}
\end{eqnarray}
Relative fluctuations in the Macquart relation are taken from the probability distribution
\begin{eqnarray}
p(\Delta_{\rm DM}) & = & A \Delta^{-\beta}_{\rm DM} \exp \left[- \frac{(\Delta^{-\alpha}_{\rm DM} - C_0)^2}{2 \alpha^2 \sigma_{\rm DM}^2} \right] \label{eq:deltadm}\\
\Delta_{\DM} & \equiv & \Delta \DMcosmic / \DMcosmic,
\end{eqnarray}
with $\alpha=3$ and $\beta=3$ as per \citet{Macquart2020}; $C_0$ is set such that the mean of the distribution is unity. This functional form has been shown to give a good description of the distribution of $\DM_{\rm cosmic}$ for a wide range of cosmological parameters and prescriptions for galactic feedback \citep{2025MNRAS.540..289G}, although the best-fit values of $\alpha$ and $\beta$ do vary. In theory, our approach could be used to fit these parameters; however, we do not consider that we have sufficient statistical power to do so, and hold these constant for the time being.

The fluctuations parameter $F$ determines the standard deviation via
\begin{eqnarray}
\sigma_{\rm DM} & = & F z^{-0.5}. \label{eq:frbf}
\end{eqnarray}
Unless otherwise noted, we fix $F=0.32$, which is consistent with the best-fit values from \citet{2024ApJ...965...57B} when using a uniform prior on $H_0$ covering both CMB and cosmological distance ladder estimates. While $F$ is a purely empirical parameter, it is inversely proportional to the baryon spread metric, which is a tracer of the strength of galactic feedback \citep{2025ApJ...983...46M}.

The final component of the DM budget, \DMhalo, is parameterised as a lognormal via $\mu_{\rm host},\sigma_{\rm host}$
\begin{eqnarray}
p({\log_{10}{\rm DM}_{\rm host}^\prime}) & = & \frac{1}{\sigma_{\rm host} \sqrt{2 \pi}} 
e^{ -\frac{(\log_{10} {\rm DM}^\prime_{\rm host}-\mu_{\rm host})^2}{2 \sigma_{\rm host}^2} }\label{eq:phost},
\end{eqnarray}
where the observed DM host is scaled from the rest-frame as
\begin{eqnarray}
    \DMhost & = & \frac{{\rm DM}^\prime_{\rm host}}{1+z}.
\end{eqnarray}
The calculation proceeds by first calculating the extragalactic dispersion measure,
\begin{eqnarray}
    \DMeg & = & \DM - \DMhalo - \DMmw,
\end{eqnarray}
and then integrating over the distributions of \DMcosmic\ (Eq.\,\ref{eq:dmcosmic} and \ref{eq:deltadm}) and \DMhost\ (Eq.~\ref{eq:phost}) to produce $P(\DMeg|z)$.

\subsection{Probabilistic calculation}

We use the z--DM code for two calculations: to find the probability of the measured radio properties, $P(\xrad) \equiv P(\DM,w,s,B)$, and the probability of redshift $z$ given those properties, $P(z|\xrad)$.

The code fundamentally calculates the rate of FRB detections as a function of DM and $z$, by first calculating the effective energy detection threshold corresponding to $s=1$, $E_0(B,w,\DM)$, giving the conditional probability
\begin{eqnarray}
R(z,\DM|B,\weff) & = & P(DM|z) \Phi(z) \frac{V(z)}{1+z} \int_{E_0}^{\infty} L(E) dE. \quad
\end{eqnarray}
The rate $R(z,\DM,B,\weff)$ is then given by
\begin{eqnarray}
  R(z,\DM,B,\weff) & = &   R(z,\DM|B,\weff) \Omega(B) P(\weff),
\end{eqnarray}
where $\Omega(B)$ is the `inverse' beamshape giving the solid angle of sky viewed at beam sensitivity $B$, and $P(\weff)$ is the distribution of widths calculated from Eq.~\ref{eq:weff}. In this work, by default, we assume $\log_{10}$-normal distributions for $w_i$ and $\tau$, which by default have $\mu_w,\sigma_w = (0,0.42)$ and $\mu_\tau,\sigma_\tau = (0.305,0.75)$ at 600\,MHz \citep{chimefrb_collaboration_first_2021}. The total rate, $R(z,\DM)$, then becomes
\begin{eqnarray}
R(z,\DM) & = & \int_{B=0}^{1} \int_{\weff=0}^{\infty} R(z,\DM,B,\weff) d\weff dB,\quad
\end{eqnarray}
Rates are related to probabilities by normalising by the total rate,
\begin{eqnarray}
R_{\rm FRB} & = & \int_0^\infty \int_0^\infty R(z,\DM) dz d\DM,
\end{eqnarray}
e.g.\
\begin{eqnarray}
    P(z,\DM,B,\weff) & = & \frac{R(z,\DM,B,\weff)}{R_{\rm FRB} }.
\end{eqnarray}
This also allows the total expected number of FRB detections can be calculated,
\begin{eqnarray}
\left< N_{\rm FRB} \right> & = & R_{\rm FRB} \Delta t,
\end{eqnarray}
given observation time $\Delta t$ (typically expressed in days). The probability of observing $N_{\rm FRB}$ is assumed to be Poissonian, such that
\begin{eqnarray}
P(N_{\rm FRB})  & = & \frac{ \left< N_{\rm FRB} \right>^{N_{\rm FRB}} e^{-\left< N_{\rm FRB} \right>}}{N_{\rm FRB}!}.
\end{eqnarray}
The probability, $P(s|z,B,w,\DM)$, can then be calculated as per Eq.~\ref{eq:ps}.

From these values, various conditional probabilities can be constructed. Of most relevance to this work is the joint probability $P(\DM,B,w,s)$, and the conditional probability $P(z|\DM,B,w,s)$. This latter is a critical input to expectations for the FRB host galaxy contribution, described in the next section.

\begin{figure}
    \centering
    \includegraphics[width=\linewidth]{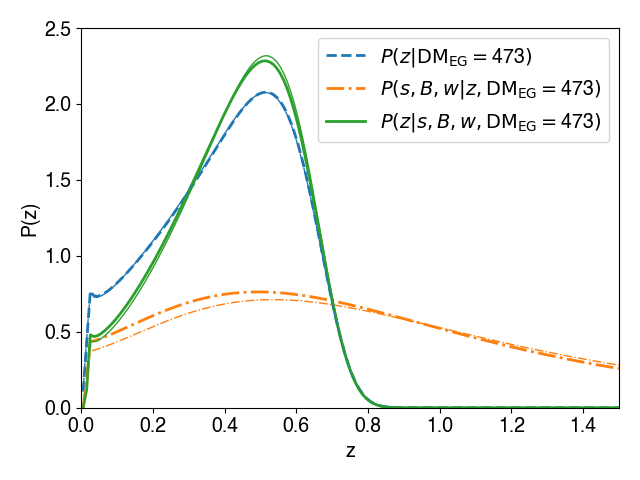}
    \caption{$P(s|z,B,w,\DM)$ for \frb, and the two components by which it is calculated: $p(z,\DM)$, and $p(s,B,w|z,DM)$, as a function of redshift. Thick lines use width and scattering distributions from \citet{chimefrb_collaboration_first_2021}; thin lines from \citet{2026PASA...43...38J}.}
    \label{fig:pz}
\end{figure}

For now, Figure~\ref{fig:pz} gives an example of the probability calculations, in the case of \frb, with $\DM=294$\,\pccc, detected by the CRAFT Incoherent Sum Survey \citep{Shannon_ICS}. We have used both the standard parameterisation of width and scattering, and the redshift-dependent prescription of \citet{2026PASA...43...38J}. Very little difference is observed between these prescriptions, justifying our approximation.

\section{Calculating $P(\xopt|\xrad)$}
\label{sec:xopt}

The primary optical properties of an image are the apparent magnitudes of identified galaxies, $m$, their angular extent, $\phi$, positions of the galaxies on sky, RA and DEC, and galaxy redshifts, $z$ (if known); secondary properties, which we do not deal with here, might include include metallicity, orientation, morphology etc. Since only the relative sky positions of candidate galaxies and the radio localisation of the FRB matter, for simplicity, we define this angular offset vector to be $\vec{x}_k$ for candidate $k$.

Our calculation of $P(\xopt|\xrad)$ proceeds according to \citet{zdm_path_hosts}, with the addition that we show how to incorporate partial redshift information. For clarity, we repeat the derivation of the method here, but extend the method to include host redshift, $z$, alongside host magnitude, $m$.

Our approach factorises the joint probability as
\begin{eqnarray}
    P(\xopt,\xrad) & = & P(\xopt|\xrad) P(\xrad) \nonumber \\
    & = & P(\xopt|U) P(U|\xrad) P(\xrad) \label{eq:joint}  \\
    && + \sum_k P(\xopt|O_k) P(O_k|\xrad) P(\xrad), \nonumber
\end{eqnarray}
where $U$ represents the instance that the true host galaxy is unseen in optical images, and $O_k$ the probability that galaxy $k$ is the true host.

\begin{figure}
    \centering
    \includegraphics[width=\linewidth]{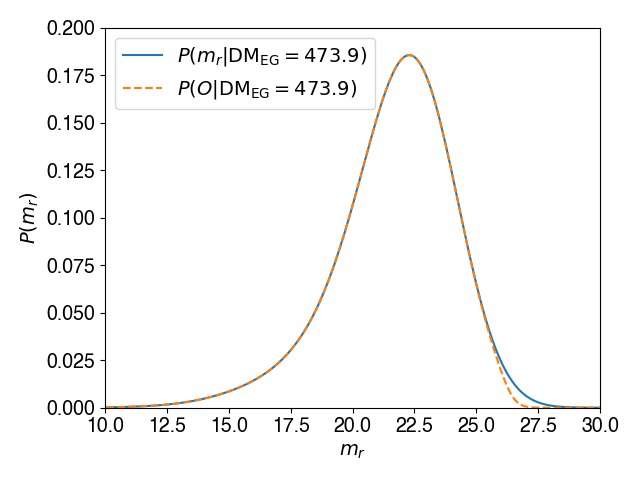}
    \caption{Prior on host galaxy magnitude $m_r$, for the case of \frb, using the Marnoch23 model. $P(O|\DMeg)$ and $P(m_r|\DMeg)$ respectively do and do not account for the chance of not identifying galaxies in an image.}
    \label{fig:pmr}
\end{figure}

To calculate $P(O|\xrad)$ and $P(U|\xrad)$, we require a host galaxy model, which can predict the FRB host galaxy magnitude $P(m|z)$. In \citet{zdm_path_hosts}, we present three such models --- we defer readers to that work for a detailed description. This is combined with a model of optical image sensitivity which defines the chance of a galaxy of magnitude $m$ being identified in the image, $P(O|m)$. Combined with the prediction for $P(z|\xrad)$ from \S~\ref{sec:xrad}, we calculate these probabilities as
\begin{eqnarray}
    P(O|\xrad) & = & \int P(O|m) P(m|z) P(z|\xrad) dz  \label{eq:po}\\
    P(U|\xrad) & = & \iint (1.-P(O|m)) P(m|z) P(z|\xrad) dz\, dm. \quad \label{eq:pu}
\end{eqnarray}
An example is given in Figure~\ref{fig:pmr}. In \citet{zdm_path_hosts}, we show how Eq.~\ref{eq:po} and \ref{eq:pu} can be used as priors in a PATH analysis, using r-band magnitudes, $m_r$. Here, we extend this to include redshift information from an arbitrary subset of candidate host galaxies.

We divide \xopt\ into the total number of galaxies observed in an image, $N_O$; their positions relative to the FRB, $\vec{x}$; and their magnitudes, $m$, and allow for $N_z$ galaxies to have their redshifts, $z$, measured, leaving $N_m$ with only magnitude information. If none of the galaxies are the true host, then ignoring galaxy clustering and treating them independently, the probability of observing this configuration is
\begin{eqnarray}
      P(\xopt|U) & = &  \Pi_{j=1}^{N_m} \rho(m_j) \Pi_{i=1}^{N_z} P_F(z_i|m_i) \rho(m_i). \quad \label{eq:unseen}
\end{eqnarray}
where $P_F$ refers to a probability relating to a field galaxy, and all $P$ refers to probabilities for the FRB host; thus $P_F(z|m)$ is the redshift distribution for field galaxies of magnitude $m$. $\rho(m)$ is the angular density of galaxies of magnitude $m$ on the sky, modified by the probability of that galaxy being detected, i.e.\ $\rho(m) =  P(O|m) \rho_{D16}(m)$, where $\rho_{D16}(m)$ is the true galaxy density, which we take from \citet{Driver2016} in the case of $r$-band magnitudes.

\begin{figure}
    \centering
    \includegraphics[width=\linewidth]{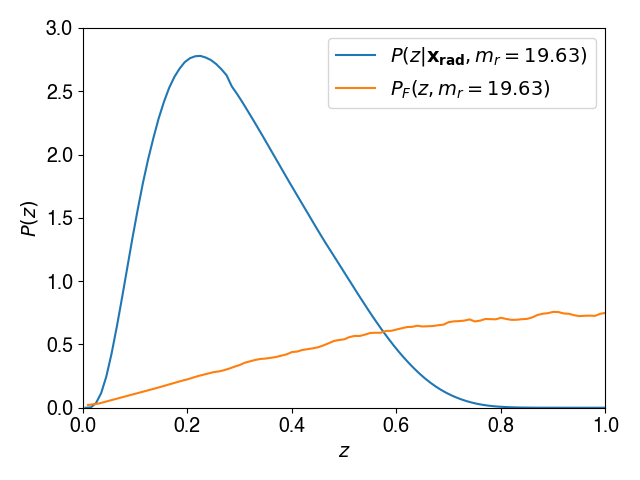}
    \caption{Probabilities of redshift $P(z|m_r)$ for both FRB hosts predicted by the Marnoch23 model, $P(z|\xrad,m_r)$, and field galaxies, $P_F(z|m_r)$, for the most probable host of \frb, with magnitude $m_r=19.63$.}
    \label{fig:pzgmr}
\end{figure}

If instead galaxy $k$ is the true host, then if $k \in N_z$,
\begin{eqnarray}
      P(\xopt|O_k \in N_z) & = & \Pi_{j=1}^{N_m} \rho(m_j) \Pi_{i=1, i \ne k}^{N_z} P_F(z_i|m_i) \rho(m_i) \nonumber \\
      && \cdot P(\vec{x}_k|O_k) P(z_k|m_k),\label{eq:pokinz}
\end{eqnarray}
and for $k \in N_m$,
\begin{eqnarray}
      P(\xopt|O_k \in N_m) & = &  \Pi_{j=1, j \ne k}^{N_m} \rho(m_j)  P(\vec{x}_k|O_k) \\
      && \cdot \Pi_{i=1}^{N_z} P_F(z_i|m_i) \rho(m_i). \nonumber \label{eq:pokinm}
\end{eqnarray}
Removing the common factor of $P(\xopt|U)$ from these equations produces
\begin{eqnarray}
\frac{P(\xopt|O_k \in N_z)}{P(\xopt|U)} & = & \frac{P(z|m_k) P(\vec{x}|O_k) }{P_F(z|m_k) \rho(m_k) }, \label{eq:pxgok1} \\
\frac{P(\xopt|O_k \in N_m)}{P(\xopt|U)} & = & \frac{ P(\vec{x}|O_k) }{\rho(m_k) }.\label{eq:pxgok2} 
\end{eqnarray}
The right-hand-sides of Eq.~\ref{eq:pxgok1} and \ref{eq:pxgok2} simply consist of odds ratios of the galaxy parameters of FRB hosts compared to field galaxies. These could be readily be extended to include parameters such as morphology, inclination, and/or metallicity. Example distributions of $P(z|\xrad,m_r)$ and $P_F(z|m_r)$ are given in Figure~\ref{fig:pzgmr}.

With these factors, we construct our full probabilities as
\begin{eqnarray}
\frac{P(\xopt,\xrad)}{P(\xopt|U)} & = & \left( \sum_{k=1}^{N_z} \frac{P(z|m_k) P(\vec{x}|O_k) }{P_F(z|m_k) \rho(m_k)} P(O_k|\xrad )  \right. \\
&& \left. +  \sum_{k=1}^{N_m} \frac{P(\vec{x}|O_k) }{\rho(m_k)}P(O_k|\xrad) + P(U|\xrad) \right) P(\xrad). \nonumber \label{eq:full_prob}
\end{eqnarray}
We note that $P(\xopt|U)$ is not a function of any model parameters, but of the distribution of field galaxies in the Universe. This varies from image-to-image, but is independent of FRB host models, population, and cosmological parameters. Hence, our calculation is somewhat simplified by ignoring it, equivalent to setting it to unity.

\begin{figure}
    \centering
    \includegraphics[width=\linewidth]{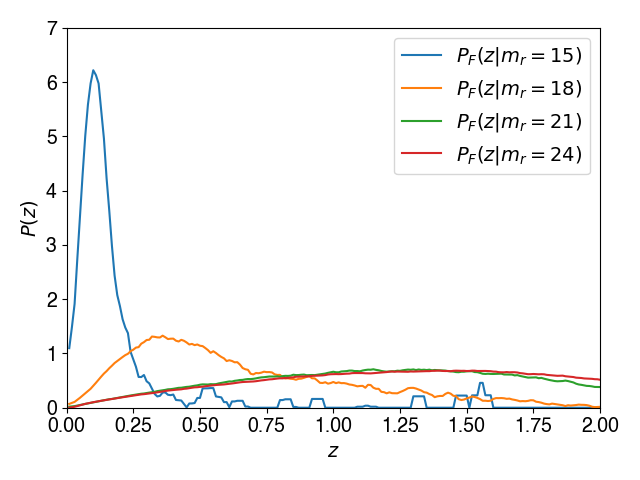}
    \caption{Example distributions of $P(z|F,m_r)$ for field galaxies (see text for method).}
    \label{fig:fieldz}
\end{figure}

We use two ingredients to calculate $P_F(z|m)$, i.e the distribution of redshift for an $m$-magnitude field galaxy. Firstly, we use a similar procedure as for the Loudas25 model from \citet{zdm_path_hosts}, where galaxies are sampled from the {\sc GALFRB} package \citep{Loudas25}. However, where the FRB host galaxy model samples galaxies weighted by their star-formation rate, SFR, or their stellar mass, $M^*$, field galaxies are sampled without weighting. This produces distributions of $P_F(m|z)$. We then calculate $P_F(z|m)$ using
\begin{eqnarray}
P(m,z) & = & P_F(m|z)P_F(z) \label{eq:pmz} \\
 P_F(z|m) & = & \frac{P_F(m,z)}{P_F(m)}.
\end{eqnarray}
We take $P_F(z)$ to be proportional to the total volume element $V(z)$ from Eq.~\ref{eq:comoving_volume}, i.e.\ assuming a constant number of galaxies per comoving volume throughout cosmological time, and calculate $P_F(m)$ by integrating $P_F(m|z)P_F(z)$ over redshift. Sample distributions of $P_F(z|m)$ are given in Figure~\ref{fig:fieldz} for different magnitudes, while the distribution for the host galaxy of FRB\,20200627 at $m_r=19.63$ is compared to $P(z|m)$ in Figure~\ref{fig:pzgmr}. Since FRB hosts tend to be dimmer than field galaxies, the probability mass is shifted towards higher redshifts in the case of field galaxies compared to true hosts. Thus, the likelihood of a host galaxy candidate being the true host will increase with decreasing redshift, as per Eq.~\ref{eq:pxgok1}.

\begin{figure}
    \centering
    \includegraphics[width=\linewidth]{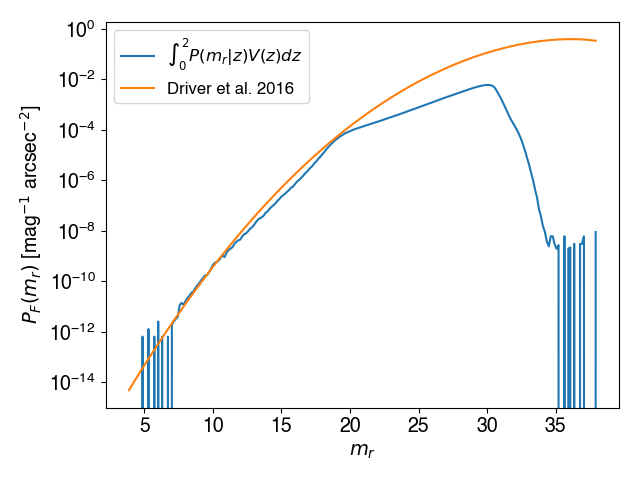}
    \caption{Comparison of r-band magnitude distributions of field galaxies, $P(m_r)$, calculated as Eq.~\ref{eq:pmz}, with the result from \citet{Driver2016}.}
    \label{fig:fieldm}
\end{figure}

To investigate the fidelity of this model, we plot the resulting $P_F(m)$ distribution in Figure~\ref{fig:fieldm}. We see a discontinuity at $z=2$, likely caused by the distributions $P_F(m_r|z)$ being calculated only up to $z=2$. 
The probability density function $\rho({\rm SFR},M^*, z)$ within {\sc GALFRB} is given by the trained neural network (NN) presented by \citet{2022ApJ...936..165L}, which is based on data from the COSMOS-2015 \citep{2016ApJS..224...24L} and 3D-HST \citep{2014ApJS..214...24S} UV-IR catalogs in the redshift range 0.2--3. In turn, this uncertainty makes the normalisation of $P_F(z|m)$ unreliable even within the $z \le 2$ range for galaxies with $m_r \gtrsim 20$, since uncertainties on the probability mass for $z>2$ affects the normalisation at all redshifts. This issue will only get worse for dimmer FRB host candidates, and we consider our current treatment to limit the accuracy of our entire model.

\subsection{Validity of partial redshift information}

Our inclusion of redshift information for a partial sample of galaxies in a field --- usually, the single most likely host, which has been selected for spectroscopic follow-up --- is valid provided that obtaining the redshift from spectroscopic information always succeeds, i.e., it is complete. If spectroscopic observations fail to identify a redshift for reasons which are correlated with the redshift itself (e.g., because the host is too faint for the spectrograph), then the resulting redshift sample becomes biased. This is a similar consideration to previous discussions about the inclusion or otherwise of redshift information within \zdm\ \citep[see e.g.][]{2025PASA...42...17H}, where partial redshift information for a subset of FRBs can be validly used provided that redshifts are always obtained for that sample. Incompleteness in redshift was treated by excluding all redshifts for FRBs with \DMeg\ above some cutoff value where the sample became incomplete. With our new methodology, we can account for this redshift incompleteness. However, should spectroscopic observations of a host candidate fail, redshift information for all host candidates at or above that galaxy magnitude should be excluded due to completeness considerations.

\section{Derived values}
\label{sec:derived}

The above section illustrates the likelihood calculation of joint FRB and optical information through Eq.~\ref{eq:full_prob}. However, there are several useful derived distributions, which we discuss below.

\subsection{Posterior host and unseen probabilities}

Our likelihood functions of Eqs.~\ref{eq:pokinz} and ~\ref{eq:pokinm} give the relative probability densities of optical images given that a galaxy is the true host, whereas Eq.~\ref{eq:pu} gives the probability given that the true host is unseen. Posterior probabilities of the true host being unseen, or any given galaxy being the true host, can therefore be calculated by normalising by the sum of these likelihoods, since these three cases (the true host is unseen, it is seen and has a measured redshift, and it is seen but does not have a measured redshift) form a mutually exclusive and collectively exhaustive set. Thus
\begin{eqnarray}
    P(U|\xopt,\xrad) & = & \frac{P(\xopt|U) P(U|\xrad) P(\xrad)}{P(\xopt,\xrad)} \\
    P(O_i|\xopt,\xrad) & = & \frac{P(\xopt|O_k) P(O_k|\xrad) P(\xrad)}{P(\xopt,\xrad)}.
\end{eqnarray}
However, inspection shows that all terms in the above include common factors of $P(\xrad)$ and $P(\xopt|U)$ (see Eq.~\ref{eq:full_prob}). For brevity, we therefore use Eq.~\ref{eq:full_prob} to write
\begin{eqnarray}
   P_{\rm norm} & \equiv & \frac{P(\xopt,\xrad)}{P(\xopt|U)  P(\xrad)} \\
   & = &  \sum_{k=1}^{N_z} \frac{P(z|m_k) P(\vec{x}|O_k) }{P_F(z|m_i) \rho(m_i)} P(O_k|\xrad )  \nonumber \\
&& +  \sum_{k=1}^{N_m} \frac{P(\vec{x}|O_k) }{\rho(m_i)}P(O_k|\xrad) + P(U|\xrad) . \nonumber \label{eq:posteriors}
\end{eqnarray}

\subsection{Probability of redshift given the true host is unseen}

\begin{figure}
    \centering
    \includegraphics[width=\linewidth]{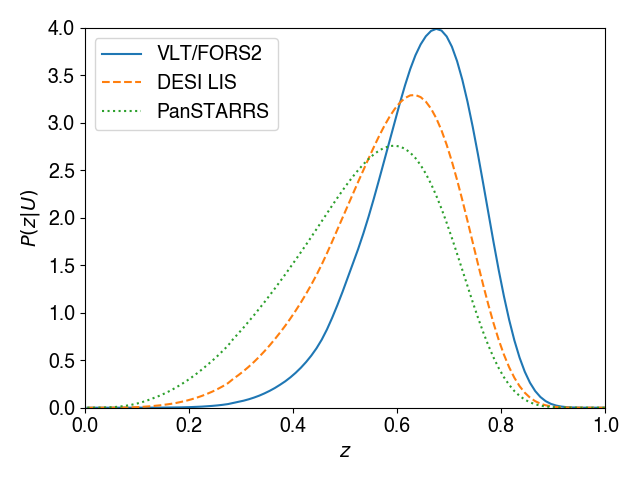}
    \caption{Probability of the redshift of \frb, given that the true host is unseen in hypothetical images taken with different instruments/surveys, with sensitivity characterised as per \citet{zdm_path_hosts}.}
    \label{fig:pzgu}
\end{figure}

Given a viable host is unseen in an optical image, the resulting redshift can be calculated as
\begin{eqnarray}
P(z|U) & = & \frac{ \int_{- \infty}^{+\infty} P(m,z|\xopt,\xrad) (1-P(O|m)) dm}{P(U|\xopt,\xrad)}.
\end{eqnarray}
This is illustrated in Figure~\ref{fig:pzgu} for the \frb, assuming three characteristic optical image depths. Clearly, the deeper an optical image, the more distant an FRB must be to be unseen in that image. However, the total redshift is still limited by the Macquart relation, such that the upper end of the $P(z|U)$ distribution suffers a similar sharp cutoff in all cases. We note that for this particular FRB, the probability $P(U)$ varies greatly between the three cases, being 1.8\% for VLT/FORS2 images, 18\% for Pan-STARRS \citep[Pan-STARRS;][]{ps1}, and 52\% for Dark Energy Spectroscopic Instrument Legacy Imaging Surveys \citep{dlis}.

\section{Demonstration of dependencies --- FRB~20190611B}
\label{sec:190611}

FRB\,20190611B was detected as part of CRAFT incoherent sum (ICS) observations, having a DM of 322.2\,\pccc\ \citep{Macquart2020}. Follow-up imaging with FORS2 in $g$, $R$, and $I$-bands revealed a host galaxy candidate (J212258.0-792350, which we denote as `\hostA') with $m_R=23.03$ located approximately $2^{\prime\prime}$ from the FRB position. The original work by \citet{Macquart2020} did not consider this galaxy to be a secure association; however, the development of the PATH framework assigned the galaxy a probability $P(O|\xopt)=0.948$ \cite{PATH}, which was later revised to $P(O|\xopt)=0.9799$ by \citet{Shannon_ICS} after minor changes to PATH itself, and a reduction in the scale size of the exponential prior to $\phi/2$. Critically, both analyses used an unseen prior $P(U)=0$.

\hostA\ has a redshift of 0.378, which is relatively high for its DM. \citet{CordesTauRedshift2022} note that its median cosmological DM estimate of $\sim$320\,\pccc\ leaves little allowance for a Milky Way or host DM contribution, especially given that FRB\,20190611B's scattering timescale of 0.18\,ms, while small, is not negligible, and thus suggestive of passage through the host interstellar medium (ISM). As noted by \citet{james_measurement_2023}, this high redshift provides a strong lower limit on the Hubble Constant, since while many structures in the Universe can produce excess DM, it is very difficult to produce large under-fluctuations (this is known as the `DM cliff' effect). Indeed, it was the very question of the fidelity of this FRB's host identification that first motivated our work on how to incorporate host uncertainty into the \zdm\ framework.

In \citet{zdm_path_hosts}, we used three models of FRB host galaxy magnitudes and a prediction for the redshift distribution from \zdm, finding an unseen prior of $P(U)=0$--$0.16$, depending on the host galaxy model being used, with a posterior confidence of 0.956--0.959 of \hostA\ being the true host. Combined with a revised high-time-resolution analysis of FRB\,20190611B giving a maximum scattering value of $0.03\pm 0.015$\,ms \citep{CRAFT_HTR}, we concluded that this galaxy is indeed the true host.

Our new framework, however, does not require any threshold for confident host classification, and also allows the redshift of \hostA\ to be used to weight its probability against the alternative of it being a field galaxy. We have also now analysed deeper optical images, which reveal another, fainter host candidate close to the FRB host localisation region. We therefore use this particular example to illustrate how our statistical method works in practice.

\subsection{Deeper optical images}

\begin{figure}
    \centering
     \begin{overpic}[width=\linewidth,trim=1cm 4cm 7.5cm 5.5cm,clip=true]{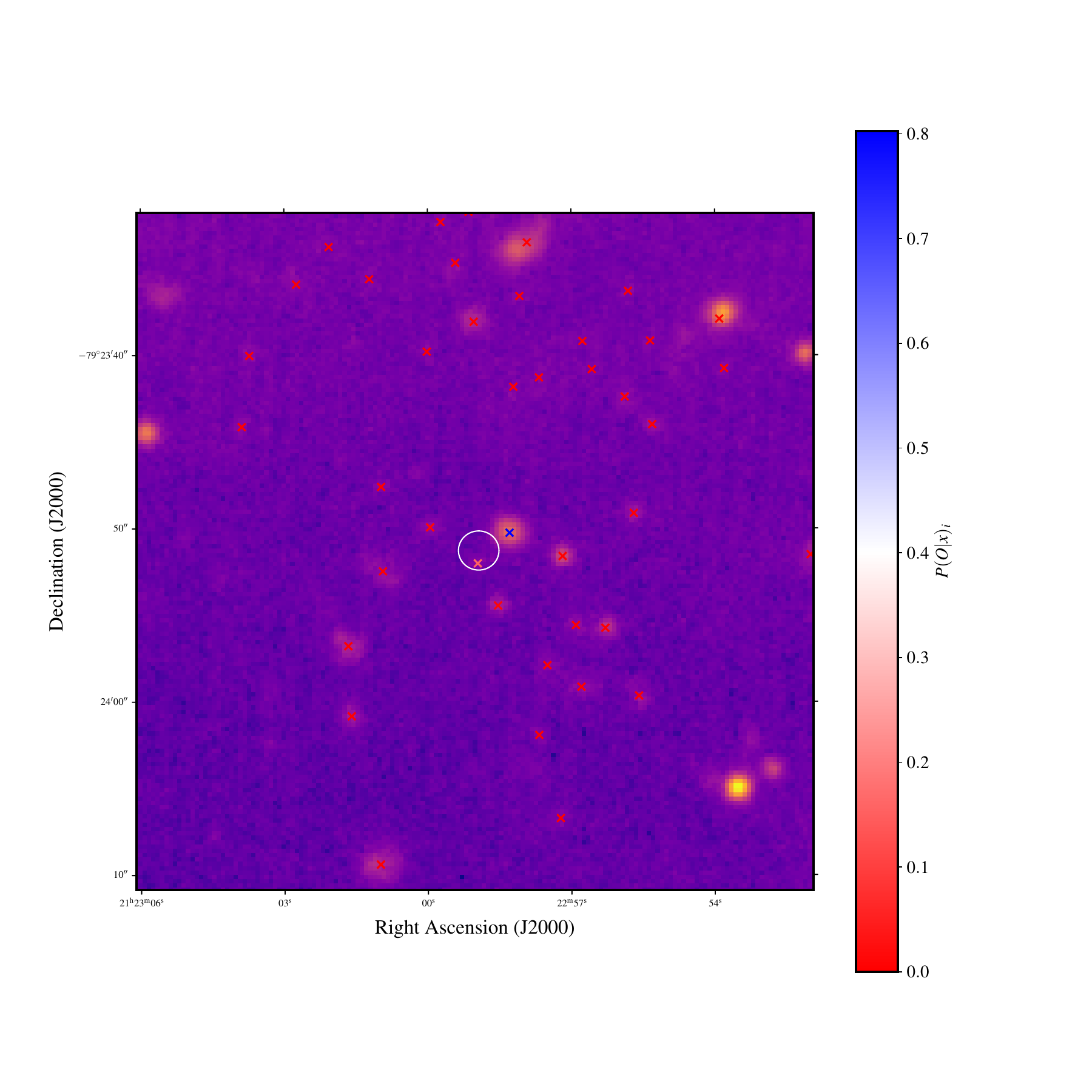}
     \put(61,54){\color{white}\textbf{A}}
     \put(54,46){\color{white}\textbf{B}}
    \end{overpic}
    \caption{R-band image of the FRB20190611B field, taken with FORS2 on the VLT. The FRB 1\,$\sigma$ localisation region is shown with a white circle, while the centroids of identified galaxies are given by crosses. Candidate galaxies A (J212258.0-792350) and B (J212259.0-792352) are marked.}
    \label{fig:190611}
\end{figure}

Motivated by the potential lack of a correct host association in \cite{Macquart2020}, several deep optical images were taken in the $R_\mathrm{special}$ filter on VLT/FORS2 on four observing blocks (OBs): 2023-11-09, 2023-11-22, 2024-05-03 and 2024-05-13. Five \SI{400}{\second} dither positions were observed for each OB (except for 2023-11-22, with one repeated position). 
These were processed using the CRAFT Optical Pipeline\footnote{\url{https://github.com/Lachimax/craft-optical-followup}}, following the procedure provided by \citet{Marnoch2023}.
To maximise the depth reached, all of the individual frames across these four epochs were co-added, for a total integration time of \SI{8400}{\second} and a depth of the combined image of approximately 26.7.
This combined image is shown in Figure~\ref{fig:190611}.
It reveals many very faint galaxy candidates not otherwise visible in the original image --- in particular, a very faint ($m_R=26.4$) host candidate J212259.0-792352 (`Galaxy B') separated by only $0.74^{\prime\prime}$ from the FRB, and within 1$\sigma$ of the localisation error. No spectroscopic observations have been taken of this candidate. We therefore apply our updated methodology to compare the relative probabilities of Galaxy A and Galaxy B being the host.

\subsection{Host analysis}

\begin{figure}
    \centering
    \includegraphics[width=\linewidth]{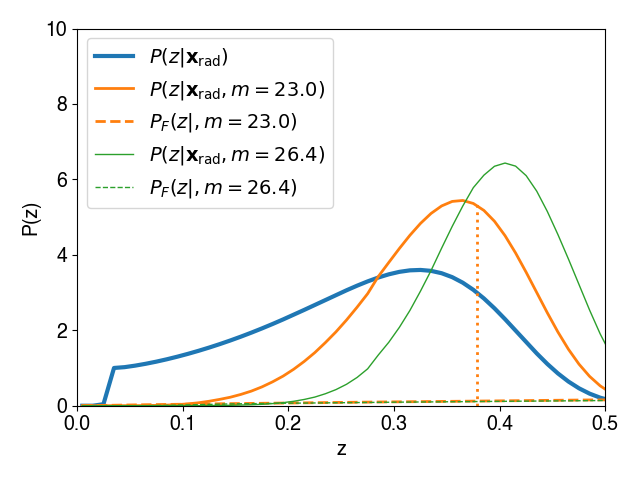}
    \caption{Predicted redshift distributions for the host galaxy of FRB\,20190611B. Shown is the predicted redshift based on observed radio properties, $P(z|\xrad)$; distributions given a specific host galaxy magnitude, $P(z|\xrad,m)$, for magnitudes corresponding the two host galaxy candidates, J212258.0-792350 (`\hostA') and J212259.0-792352 (`\hostB'); and the predicted redshift distributions of these galaxies if they are field galaxies, $P_F(z|m)$. The detected redshift of \hostA, $z=0.378$, is also indicated.}
    \label{fig:190611_pzgdm}
\end{figure}

We fix the standard \zdm\ parameters from \citet{HoffmannHalo}, which gives the $P(z|\xrad)$ distribution given in Figure~\ref{fig:190611_pzgdm}. We see that the predicted redshift distribution based on radio properties, $P(z|\xrad)$, has a peak near $z=0.325$, and cuts off by $z=0.5$ --- the redshift of 0.378 of \hostA\ is well below the cutoff at $z=0.5$, and agrees very closely with the predicted redshift distribution once its magnitude is considered, $P(z|\xrad, m=23.03)$. However, if \hostB\ is the true host, its expected redshift, $P(z|\xrad, m=23.03)$, is likely to be even higher than that of \hostA, and hence be more discrepant from the Macquart Relation. Furthermore, this would require \hostA\ to be a field galaxy of very low redshift for its faint magnitude, since the probability distribution, $P_F(z|m=23.03)$, peaks at $z>1$. Hence, it is far more likely that \hostA\ is the true host.

\begin{table*}[]
    \centering
    \begin{tabular}{r|ccccccccc}
    Candidate     & $m_r$ & $\theta$ & $\phi$ & $P(O|\xrad)$ & $\rho(m)$ & $P(\vec{x}|O)$ & $P(z|m)$ & $P_F(z|m)$ & $P(O|\xrad,\xopt)$ \\
Galaxy A: J212258.0-792350  & 23.03 & 2.05 & 0.54 & 0.127  & 0.128  & 0.0013 & 4.95 & 0.32 & 0.97\\
Galaxy B: J212259.0-792352              & 26.4  & 0.74 & 0.29 & 0.00029 & 0.020085 & 0.020 & N/A & N/A & 0.0006\\
    \end{tabular}
    \caption{Calculated probabilities of hosts for FRB\,20190611B, showing the relative posterior likelihoods for the original host galaxy, 
    J212258.0-792350 (`\hostA'), against that of the alternative, J212259.0-792352 (`\hostB'), using our combined \pth\ and \zdm\ framework.}
    \label{tab:190611}
\end{table*}

In Table~\ref{tab:190611}, we give the relevant likelihood values for optical parameters (the relative likelihood of this FRB being detected at all, $P(\xrad)$, is independent of the host scenario). While the probability $P(\vec{x}|O)$ of observing the FRB localisation ellipse given that a galaxy is the true host is four times more likely for the fainter galaxy, both the magnitude (relative values of $P(O|\xrad)$ and $\rho(m)$), and the redshift (relative values of $P(z|m,\xrad)$ and $P_F(z|m,\xrad)$, point to \hostA\ being the true host, with a probability $P(O|\xrad,\xopt) = 0.97$.

We also note that the density of $m=26.4$ galaxies on the sky is approximately one per $50$\,arcsec$^2$ --- it is therefore not surprising that, in a sample of 35 localised FRBs, one has such a dim field galaxy overlapping its localisation region. It would be extremely surprising, however, were Galaxy B to be the true host, that it would coincidentally be located on the sky very close to a rarer $m=23.03$ galaxy --- especially one with such a low redshift. Both considerations are quantitatively included in our new framework, and the resulting likelihoods reflect these arguments.

\section{Verification}
\label{sec:sandbox}


\begin{figure}
    \centering
    \includegraphics[width=\linewidth]{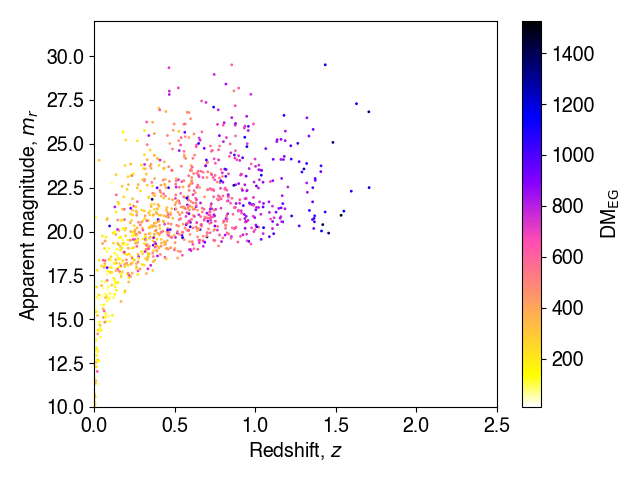}
    \caption{Distribution of Monte-Carlo sampled FRBs and host galaxies from the ASKAP/CRACO 900\,MHz survey.}
    \label{fig:craco_scatter}
\end{figure}

We now proceed to show that our methodology correctly reproduces model parameters. We model the CRAFT CRACO system \citep{2025PASA...42....5W}, with survey parameters from the 900\,MHz band at 12.8\,ms time resolution \wangjames, using the FRB population and cosmological parameters from \citet{HoffmannHalo}.
We then sample Monte-Carlo-generated FRBs as performed in \citet{2022MNRAS.516.4862J}, producing the set $ \{ z, DM, w, B, s \}$. The apparent magnitude distribution of host galaxies is taken from the `Loudas25' model described in \citet{zdm_path_hosts}, and based on the work of \citet{Loudas25}. We choose this model because it allows model fitting by varying a single parameter, $f_{\rm sfr}$. For this test, we set $f_{\rm sfr}=1.5$, approximately the lower limit found by \citet{zdm_path_hosts}, reflecting that FRB host galaxies tend to be fainter than predicted when tracing star-formation, and much fainter than predicted by following stellar mass.

The resulting sample of 1,000 FRBs is shown in the scatter plots of Figure~\ref{fig:craco_scatter}.

\begin{figure}
    \centering
    \includegraphics[width=\linewidth]{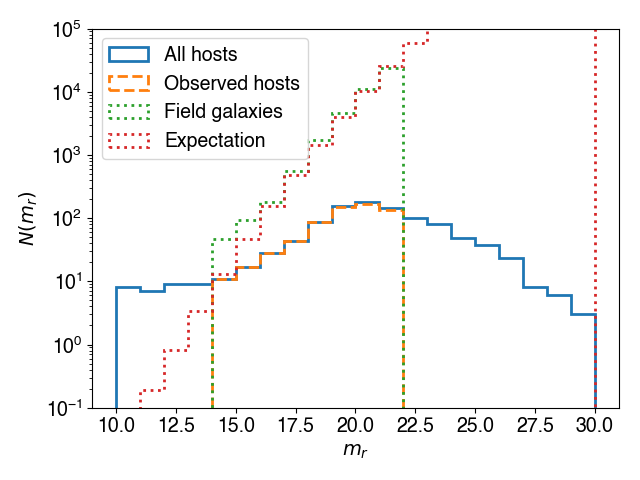}
    \caption{Histogram of galaxy magnitudes, representing the true host magnitude distribution estimated by \zdm; the visible host distribution of matched galaxies taken from the DESI Legacy surveys; the distribution of non-host galaxies in the $2'$ cutouts about the simulated galaxy locations from the $30"$ localisation sample; and the expected field galaxy angular density from \citet{Driver2016}.}
    \label{fig:hostmaghist}
\end{figure}

Each of these simulated hosts are assigned a real host galaxy according to the procedure of \Andersen. Briefly, each FRB's nominal host $m_r$ is matched to a real galaxy detected in the DESI Legacy Imaging Surveys \citep{dlis} for $14 \le m_r \le 22$, the 
the Hyper Suprime-Cam Subaru Strategic Program (HSC-SSP) $3^{\rm rd}$ public data release for $22 < m_r$ \citep{2022PASJ...74..247A}, and the Heraklion Extragalactic Catalogue \citep[HECATE;][]{2021MNRAS.506.1896K} for very bright hosts ($m_r < 14$). We assume a true FRB offset distribution of an exponential with scale size $\phi$ equal to half of the host's angular scale $\psi$, and generate a random offset corresponding to a localisation uncertainty of $\sigma_{\rm FRB}=0.5"$. Cutouts of $10"$ square are then taken about each nominal FRB position using the DESI Legacy surveys only. A comparison of the (simulated) distributions of galaxy magnitudes for all hosts, visible hosts, and ``field'' galaxies (i.e., catalogue galaxies within the $10^"$ cutouts that are not hosts) is given in Figure~\ref{fig:hostmaghist}. Since FRB host galaxies were selected from this survey in the magnitude range $14 \le m_r \le 22$, we also restrict the field galaxies the same range. This leaves 623 true FRB hosts, 377 FRBs with no true host, and 440 field galaxies in the sample.

As noted in the Introduction, FRB observations have a wide range of localisation fidelities. To investigate how our procedure behaves in a regime where localisation is relatively poor, we re-run the galaxy assignment procedure of \Andersen\ when using an accuracy of $\sigma_{\rm FRB}=30"$, and extend the region of analysed galaxies to be all within $2'$ of the simulated FRB position. This means that the host offset distribution $f(\theta/\phi)$, and hence $P(O|\vec{x})$, contains no information, since all except the most nearby galaxies will be point-like with respect to the localisation. In this sample, the 41,800 field galaxies dominate over the 637 hosts.

We can see from Figure~\ref{fig:hostmaghist} that the angular density of field galaxies in the DECaLS catalogue matches the expectation from \citet{Driver2016}, when the $30"$ localisation sample is analysed. However, we see a slight excess of field galaxies beyond expectations at low magnitudes.
We therefore derive a correction function $c(m_r)$ of the form
\begin{eqnarray}
    \rho^\prime & = & c(m_r) \rho_{D16}  \nonumber \\
    c(m_r) & = & 1.03+4.26 \exp(14.0- m_r), \label{eq:rho_correction}
\end{eqnarray}
and use it where noted below.

\begin{table}[]
    \centering
\begin{tabular}{l|cc|c|cc}
\hline
& \multicolumn{2}{c|}{Priors} & & \multicolumn{2}{c}{Estimates} \\
Parameter & Min & Max & Truth & Combined & Confidant\\
\hline 
$\log_{10} F$ & -2 & 0 & -0.49 & $-0.64_{-0.16}^{+0.19}$ & $-0.52_{-0.21}^{+0.21}$\\
\nsfr & 0.0 & 4.0 & 2.88 & $3.05_{-0.94}^{+0.65}$ & $2.9_{-1.1}^{+0.8}$\\
$\alpha$ & -4 & 0. & -1.55 & $-2.2_{-1.6}^{+2.3}$ & $-2.7_{-2.0}^{+2.6}$\\
\muhost & 1.0 & 3.0 & 2.13 & $2.16_{-0.09}^{+0.09}$ & $2.17_{-0.09}^{+0.09}$\\
\sighost & 0.1 & 1.5 & 0.46 & $-0.52_{-0.08}^{+0.08}$ & $0.46_{-0.08}^{+0.10}$\\
\lemax & 40 & 43.0 & 40.9 & $40.84_{-0.08}^{+0.10}$ & $40.91_{-0.12}^{+0.12}$\\
$\gamma$ & -2.0 & 0.0 & -1.12 & $-1.08_{-0.05}^{+0.06}$ & $-1.07_{-0.06}^{+0.06}$\\
$H_0$ [\pccc] & 60 & 80 & 70.63 & $75.0_{-7.2}^{+6.7}$ & $70.7_{-7.8}^{+8.5}$\\
\DMhalo & 0 & 100 & 68 &  $67.0_{-9.6}^{+7.0}$ & $61.2_{-14.0}^{+9.4} $\\
$f_{\rm sfr}$ & 0 & 2 & 1.5 & $1.57_{-0.07}^{+0.06}$ & N/A  \\
$\phi$ & 0.01 & 6 & 0.5 & $0.45_{-0.03}^{+0.03}$  & N/A\\
\hline
\end{tabular}
    \caption{Parameters fit to our synthetic FRB data, giving the prior minimum and maximum ranges, truth values, and estimates using combined \pth\ and \zdm\ analysis developed in this work, and using only confidant hosts with $P(O|X)>0.95$ in a \zdm-only analysis. Parameters are: $F$, the `fluctuation' parameter from Eq.~\ref{eq:frbf} governing the intrinsic variation in \DMcosmic; $\nsfr$, which modulates the scaling of the FRB population with the star-formation rate according to Eq.~\ref{eq:nsfr}; $\alpha$, being the frequency-dependent FRB rate in Eq.~\ref{eq:rate_scaling}; $\mu_{\mathrm{host}}$ and $\sigma_{\mathrm{host}}$ are the mean and standard deviation of the assumed log-normal distribution of host galaxy DMs; \emax, and $\gamma$ are the parameters of the FRB luminosity function, which is parameterised as a Schechter function via Eq.~\ref{eq:schechter}, with $\lemin=30$; $H_0$ is the Hubble constant; $f_{\rm sfr}$ is the scaling of the FRB host galaxy magnitudes with the star-formation rate according to the Loudas25 model; and $\phi$ is the exponential scale size of FRB offsets from their host galaxy centres relative to the galaxy's half-light radius.}
    \label{tab:params}
\end{table}

\subsection{MCMC verification}

In order to show the utility of our likelihood function of Eq.~\ref{eq:full_prob}, we run an MCMC using Python's {\sc emcee} package \citep{emcee} as per \citet{2025PASA...42...17H}.  We allow our full set of \zdm\ parameters to vary: the Hubble constant $H_0$, fluctuation parameter $F$, mean $\mu_{\rm host}$ and standard deviation $\sigma_{\rm host}$ of $\log_{10} {\rm DM}_{\rm host}$, the Milky Way halo contribution ${\rm DM}_{\rm halo}$, and FRB energy function parameters $\log_{10} E_{\rm max}$ and $\gamma$, respectively describing the turnover energy (in ergs, assuming 1\,GHz emission bandwidth) and cumulative slope of the FRB energy function; scaling of the FRB rate according to the star-formation rate to the power $n_{\rm sfr}$; and $\alpha$, which governs the frequency-dependence of the FRB rate as $R \sim \nu^\alpha$. We do not fit $E_{\rm min}$ in this dataset, due to numerical instabilities regarding the sharp minimum cut-off in the energy function, as described in Appendix~\ref{sec:emin}.

We fit the data using the Loudas25 model, with free parameter $f_{\rm sfr}$, and also fit for the exponential width of the FRB offset distribution, $\theta_0$. These parameters, their simulated truth values, and their (uniform) priors, are defined in Table~\ref{tab:params}.

\begin{figure*}
    \centering
    \includegraphics[width=\textwidth]{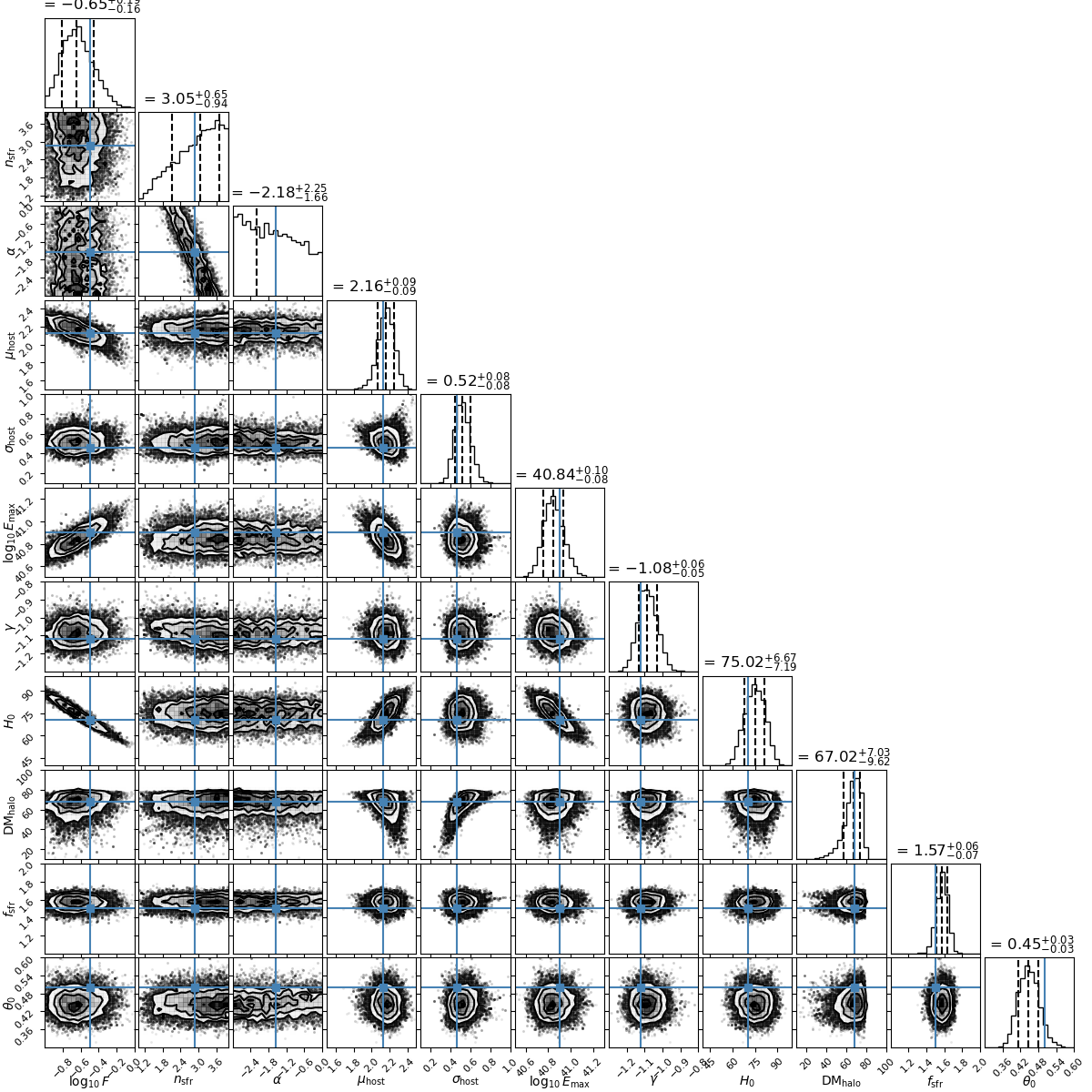}
    \caption{Cornerplot showing parameter estimates based on our synthetic dataset. In the single-parameter plots, simulated truth values are shown as vertical blue dashed lines, while vertical black dashed lines indicate the 16\%, 50\%, and 0.84\% quantiles.}
    \label{fig:bridgetcorner}
\end{figure*}

\begin{figure}
    \centering
    \includegraphics[width=\linewidth]{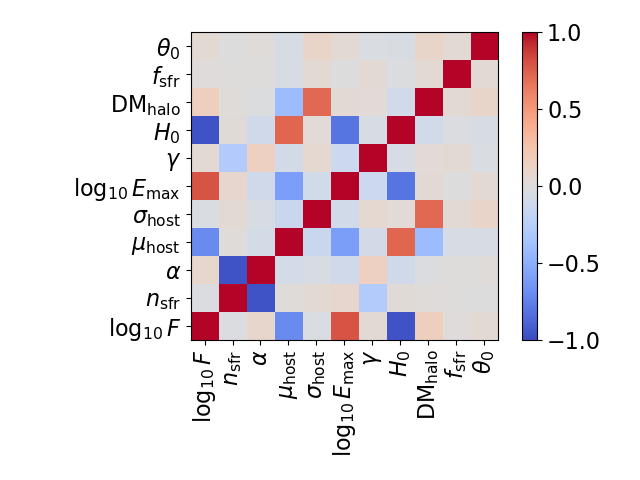}
    \caption{Pearson correlation coefficients between our estimated parameters, using the synthetic sample of CRACO FRBs generated according to the methods of \Andersen.}
    \label{fig:bridgetcorrelation}
\end{figure}

We estimate the likelihood according to the equations given in \S~\ref{sec:xopt}, and use the {\sc emcee} \citep{emcee} package to run a Markov Chain Monte Carlo with 40 walkers, with 2400 steps each. We allow for a burn-in of 1000 steps per walker, giving $40\times 1400$ total samples. The resulting corner plot and parameter estimates are given in Figure~\ref{fig:bridgetcorner} and Table~\ref{tab:params}, respectively. Pearson correlation coefficients between parameters are given in Figure~\ref{fig:bridgetcorrelation}.

We see from Figure~\ref{fig:bridgetcorner} that all parameters are correctly estimated (i.e., the simulated truth lies within a small, well-defined error range), and deviations from the expected mean values are purely due to random statistical errors, with two exceptions. The first is $\alpha$, where a single survey contains very little constraining power --- primarily due to its degeneracy with $n_{\rm sfr}$. Thus constraints are primarily due to the priors. The other is $\theta_0$, the characteristic scale size of the exponential distribution of FRB offsets from their host centres. We show in \S\,\ref{sec:image_size} that this is due to the $10" \times 10"$ image size missing relatively nearby hosts.

As shown in Figure~\ref{fig:bridgetcorrelation}, our analyses reproduces well-studied correlations in FRB population parameters, i.e.\ between $H_0$, \emax, \muhost, and $\log_{10} F$ \citep{2024ApJ...965...57B};
\DMhalo, \muhost, and \sighost\ \citep{HoffmannHalo}; and $\alpha$ and $n_{\rm sfr}$ \citep{james_zdm_2022}. Importantly, we find only very weak correlations between our optical parameters ($f_{\rm sfr}, \theta_0$) and others. 
Nonetheless, the fact that weak correlations do exist (e.g.\ -6\% between $\theta_0$ and $H_0$, or 8\% between $f_{\rm sfr}$ and \DMhalo) illustrates that precision cosmological measurements will require an accurate estimate of FRB host behaviour, to avoid the influence of host galaxy bias.

\subsubsection{Biasing effects of image size}
\label{sec:image_size}

\begin{figure}
    \centering
    \includegraphics[width=\linewidth]{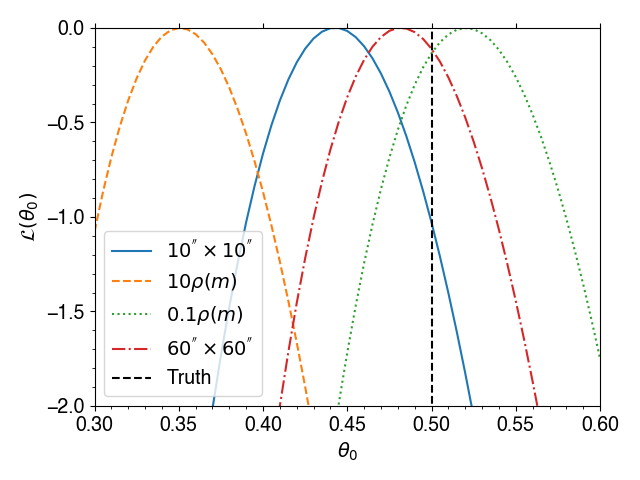}
    \caption{Likelihood of exponential scale size of FRB offset from host galaxy centre, $\mathcal{L}(\theta_0)$, evaluated on our simulated CRACO sample of FRB host galaxies. Shown are estimates for the standard sample of 1,000 FRBs, using images of simulated size $10{"} \times 10{"}$, with a systematic bias towards low values of $\theta_0$;  evaluations of 1,000 FRBs where the background galaxy density $\rho(m)$ has been artificially increased to $10 \rho(m)$ and decreased to $0.1 \rho(m)$; and when using simulated images of $60{"} \times 60{"}$.}
    \label{fig:theta0}
\end{figure}

Our MCMC procedure estimates $\theta_0 = 0.45\pm 0.03$, compared to a simulated truth of $0.5$. In Figure~\ref{fig:theta0}, we plot $\mathcal{L}(\theta_0)$ when holding all other parameters constant at their simulated truth values, reproducing the bias with a best-estimate of $\theta_0=0.44$.

What causes it? An incorrect estimate of the field galaxy density $\rho(m)$ affects $\theta_0$ through Eq.~\ref{eq:full_prob}. Increasing $\rho(m)$ decreases the factor multiplying $P(\vec{x}|O)$ for field galaxies, which causes the likelihood to maximise by decreasing $P(\vec{x}|O)$ further. For field galaxies which are (on average) further from the FRB, this means decreasing $\theta_0$.  As shown in Figure~\ref{fig:theta0}, artificially increasing $\rho(m)$ by a factor of 10 causes the estimated $\theta_0$ to decrease by $\sim$0.1, and vice versa. However, while our estimates of background density are imprecise (see Figure~\ref{fig:hostmaghist} and Eq.~\ref{eq:rho_correction}), they are not sufficiently inaccurate to account for this effect.

It turns out that this bias is due to the $10" \times 10"$ image size used in our synthetic sample, which fails to include the host for FRBs located in the outskirts of large, nearby galaxies. For our simulated sample, galaxies with $14 \le m_r le 15$ have a mean half-light radius of $8.6"$, meaning that these images will miss a large fraction of all true hosts --- and hence, host offsets. These missing, high-offset hosts will cause the estimated offset distribution to be biased towards FRBs close to their host centres, i.e., a low value of $\theta_0$. Using simulated images of $60'' \times 60''$ causes the estimated value of $\theta_0$ to increase from 0.45 to 0.48. We therefore conclude that this is the cause of the observed bias.

\subsection{Verification of posterior distributions}

\begin{figure}
    \centering
    \includegraphics[width=\linewidth]{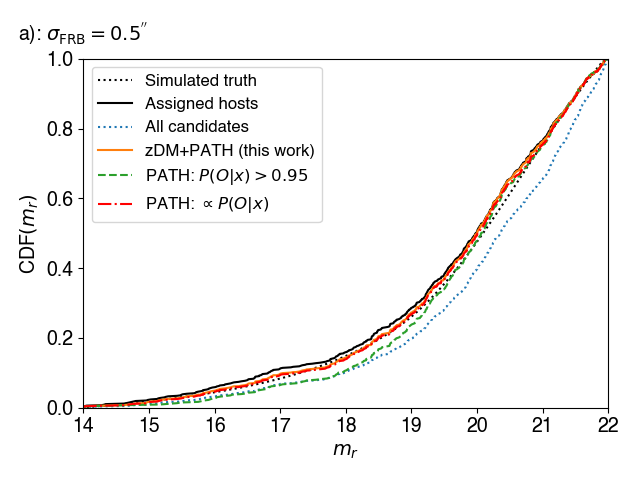}
    \includegraphics[width=\linewidth]{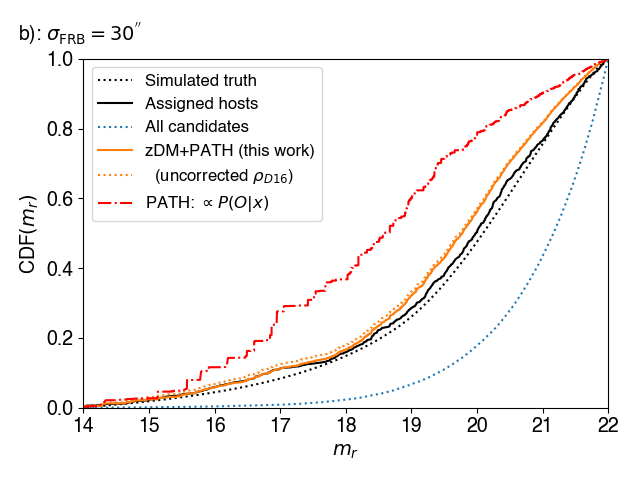}
    \caption{Cumulative magnitude distributions of FRB host galaxy candidates for the synthetic samples with localisation accuracy of a: $\sigma_{\rm FRB} = 0.5^{"}$, and b: $30^{"}$. Shown are the true host galaxy distribution; distribution of assigned hosts according to the procedure of \Andersen; the posterior distribution found via the combined \zdm\ and \pth\ analysis of this work; the same, but for an uncorrected galaxy density function; the resulting posterior distributions when performing a traditional PATH analysis by assuming $P(U)=0.1$ and taking only `firm' candidates with $P(O|x)>0.95$; and the same, but without cutting on $P(O|x)$, and instead weighting all candidates by $P(O|x)$. Note that in a), the \zdm\ + \pth\ and \pth\ $\propto P(O|x)$ curves overlap, while in b), no candidates satisfied the $P(O|x)>0.95$ criteria.}
    \label{fig:cum_mags}
\end{figure}

To more precisely test the utility of our framework in reproducing the underlying FRB host galaxy distributions, we compare in
Figure~\ref{fig:cum_mags} the cumulative magnitude distributions of the underlying truth (both the model, and the distribution of hosts assigned by the procedure of \Andersen), and the posterior distributions when weighting candidates by $P(O|\xopt,\xrad)$ from Eq.~\ref{eq:posteriors}.
We also compare this against results from a \pth\ only analysis when setting priors $P(U)=0.1$. In this latter case, we construct posterior distributions both by selecting `confident' associations only, i.e.\ only host candidates with $P(O|\mathbf{x})>0.95$, and when weighting all candidates by their posteriors, $P(O|\mathbf{x})$. This is repeated for both the $\sigma_{\rm FRB}=0.5"$ and $\sigma_{\rm FRB}=30"$ localisation samples.

Applied to the $\sigma_{\rm FRB}=0.5"$ sample, we see that both our combined \zdm\ and \pth\ framework, and the \pth-only analysis when weighted by $P(O|x)$, reproduce the distribution of assigned hosts to very high accuracy. The correction factor to $\rho_{D16}$ has negligible effect, since in this sample, the term $P(x|O)$ (for the probability of observing an FRB at the given position $x$ given $O$ is the host) in Eq.~\ref{eq:full_prob} contains more information than the relative number of candidates compared to expectations, $\rho(m)$. Using confident hosts only (of which there are 537) produces a slight bias against bright host galaxies, which get down-weighted compared to the distribution of all host candidates. We suspect this is because the large angular size of bright galaxies limits the maximum value of $P(x|O)$, since that probability gets distributed over a large region of sky --- in the limit that a galaxy takes up the entire sky, the term $P(x|O)$ will be $(4 \pi)^{-1}$\,sr$^{-1}$. This in turn limits the maximum value of $P(O|x)$ for these galaxies, pushing them below the $0.95$ `confidant' threshold.

Applied to the $\sigma_{\rm FRB}=30"$ sample, our combined \zdm\ and \pth\ methodology well-reproduces the true host distribution, albeit with a slight excess of $18 < m_r < 21$ hosts. We expect that this difference is statistically significant, and may arise from within either the \zdm\ or \pth framework. It is not due to the correction factor to the background galaxy density $\rho(m_r)$ of Eq.~\ref{eq:rho_correction}, which produces only a small change in the distribution. We leave studying this systematic to a future analysis, perhaps similar to the methods of \Andersen, to uncover this error. What is certain however is that our estimates produce a vast improvement over the a \pth-only analysis: not a single candidate passes the $P(O|x)>0.95$ confidence cut, and even a \pth\ analysis which has been correctly weighted by $P(O|x)$ produces a highly biased posterior distribution towards bright galaxies.

\subsection{Estimation of host galaxy bias}
\label{sec:bias_estimate}

To estimate the potential bias of only using well-identified FRB hosts in population analyses, we construct the following example. We take only the 537 FRBs from our synthetic $\sigma_{\rm FRB}=0.5"$ sample identified to have $P(U|x) > 0.95$ in a traditional PATH analysis as described above, and use these in a \zdm-only analysis studying FRB population properties. This very approximately simulates the sample of FRBs used by \citet{DSA_Sharma_sfr} and \citet{2025NatAs...9.1226C}, which primarily uses optical data from the DESI Legacy Imaging Surveys \citep{dlis} to identify FRBs detected by DSA, which has a sensitivity similar to that of CRACO. To avoid artificial bias due to our $m_r \ge 14$ cut, we assume all FRBs with true hosts of magnitude $m_r < 14$ have a correctly identified host.

The resulting parameter estimates are compared to truth values in Table~\ref{tab:params}. Differences in parameter estimates between this and the combined \pth\ and \zdm\ analysis above are due both to biases from using confidant galaxies only, and also random statistical errors from the properties of the 463 FRBs without confidant hosts (since precisely the same sample of FRBs is used, random errors from FRBs with confidant hosts behave identically). This shows that any biases present when using confidant host galaxies only are smaller than random statistical errors, since approximately half the parameters ($F,\nsfr,\sighost,\emax,H_0$) are better-estimated with confidant hosts only, and the other half ($\alpha,\muhost,\gamma,\DMhalo$) are not. This suggests that these effects are not very important when using a sample of 1000\,FRBs.

We observe however that parameter estimates made using the combined analysis are more precise (reduced errors) for some parameters ($\nsfr,\alpha,\sighost,\emax,H_0,\DMhalo$; error reductions of $\sim$20\%), while they are equally precise for others ($F,\muhost,\gamma$). This is due to the information contained in the FRBs without confidant hosts --- both through their DM, and constraints placed on their redshift.

This suggests that the primary advantages of using our updated method are an unbiased distribution of FRB hosts, and increased precision in host identification, with a small increase in precision of FRB population parameter estimates being a secondary benefit.

\section{Analysis of CRAFT data}
\label{sec:craft}

\begin{figure*}
    \centering
    \includegraphics[width=\textwidth]{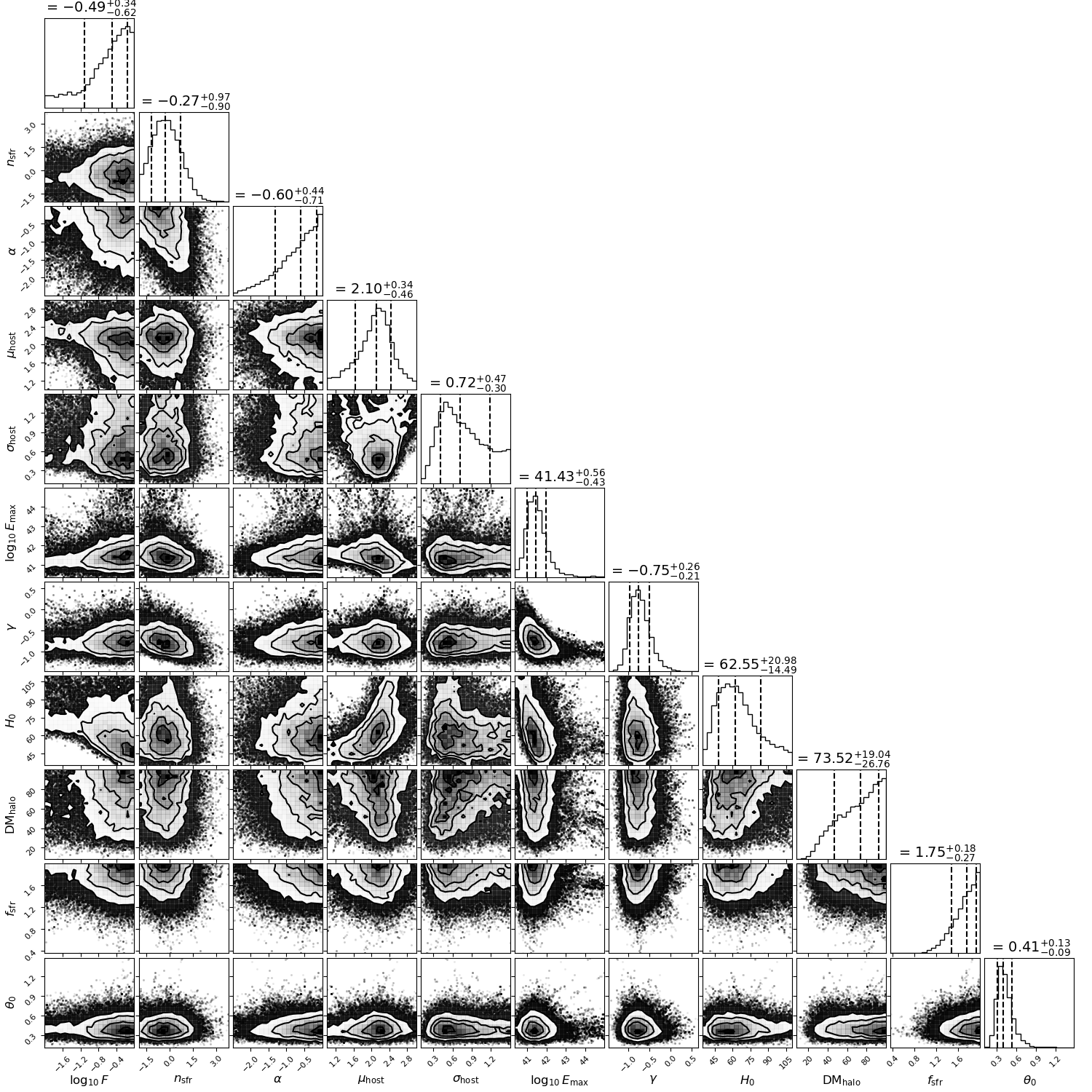}
    \caption{Cornerplot of our MCMC analysis of the CRAFT ICS sample. Definitions of parameters are given in Table~\ref{tab:params}. See text for explanations.}
    \label{fig:ICS_cornerplot}
\end{figure*}

Having verified the ability of our model to reproduce simulated truths in the face of experimental biases, we proceed to apply it to real FRB data.

We now apply our methodology to the ASKAP/CRAFT ICS survey, which are modelled in \zdm\ as described in \citet{james_measurement_2023}. This dataset has now been expanded by \citet{HoffmannHalo} to consist of 43 FRBs presented by \citet{Shannon_ICS}, detected over the frequency range 663.5--1799.5\,MHz, up until mid 2024. Of these, 37 have ${\rm DM},s,B,\weff \in \xrad$, and six FRBs have ${\rm DM},s,\weff \in \xrad$. This is because the beam values $B$ are taken from \citet{2025PASA...42....3A}, which used a subset of these FRBs up until the end of 2023. While the scattering $\tau$ and intrinsic widths $w_i$ of these bursts have been analysed in \citet{CRAFT_HTR}, and can be used to constrain the FRB scattering and intrinsic width distributions as shown by \citet{2026PASA...43...38J}, only $\weff$ affects detectability. Modelling these intrinsic distributions allows for their redshift-dependence, and can affect the inferred $z$-dependent rate. However, there is no current empirical evidence for a redshift dependence of these properties, and including such dependence in \zdm\ causes the code to take significantly longer to run, so we do not use them here.

Optical properties $\xopt$, i.e.\ galaxy magnitudes, positions, and angular sizes, are available from \citet{Shannon_ICS} for a total of 32 of these FRBs ---
note that the image of FRB~20240210A was accidentally omitted from this work, but has since been published by \citet{2025ApJ...993..119G}. To these we add the FRB~20210912A field, which as shown by \citet{Marnoch2023}, did not yield any apparent host. All images are taken by the FOcal Reducer and low dispersion Spectrograph \citep[FORS2; ][]{FORS2} on the ESO Very Large Telescope (VLT) in a combination of $g$, $R$, and $I$ bands --- we approximately adjust between bands according to \citet{zdm_path_hosts}, by adding/subtracting 0.65 from $I$ and $g$ band magnitudes, respectively, to scale to $R$ band magnitudes. The probability of identifying a galaxy within these images, $P(O|m)$, is also modelled in \citet{zdm_path_hosts}, where $P(O|m)$ is 50\% at $m_R=26.18$, and is 90\% (10\%) at 25.42 (26.93).
This makes for a total of $33$\,FRBs for which $P(\xopt,\xrad)$ is calculated according to Eq.~\ref{eq:full_prob}, and 10 for which only $P(\xrad)$ is calculated.

We optimise over the same set of parameters given in \S\,\ref{sec:sandbox}, using an MCMC with 40 walkers, and 6000 steps, again discarding the first 1,000 as a burn-in sample. The results are shown in Figure~\ref{fig:ICS_cornerplot}.

Firstly, we note that the limits shown on cosmological ($F$, $H_0$) and FRB population parameters (e.g., $\alpha$) are not the most stringent in the literature. The CRAFT ICS sample spans a limited range of frequency and sensitivity, and contains relatively few high-redshift FRBs.
In particular, the measured rates are inconsistent with previous observations in Fly's Eye mode \citep{2019PASA...36....9J,Shannon_ICS}, making estimation of $\alpha$ (which is mostly influenced by the measured rate as a function of observing frequency) particularly poor.  Therefore, we find relatively weak constraints on most parameters, and several are not constrained within their prior limits. 
Combining these observations with localised FRBs from more-sensitive telescopes, such as MeerTRAP \citep{PastorMarazuela2025} and DSA \citep{DSA_Sharma_sfr}, gives the most up-to-date limits on these parameters --- see \citet{HoffmannHalo} for our most recent results.

Hence, in this calculation, we are not overly concerned with parameters pressing up against their priors. Rather, we seek to show that our extended likelihood analysis does not skew previously well-established results with this dataset --- and our results are clearly consistent with those of \citet{HoffmannHalo}.

Our results also reproduce the strong preference for high values of $f_{\rm sfr} > 1$ originally seen in \citet{zdm_path_hosts}, which as discussed there, is driven by the FRB hosts being predominantly faint galaxies. We also quantitatively constrain the exponential offset distribution scale parameter $\theta_0 = 0.41_{-0.09}^{+0.13}$, which is the first time this parameter has been fit while considering the influence of field galaxies. Our value is consistent with $\theta_0=0.5$ found by \citet{Shannon_ICS} when assuming confident host associations. We leave it to other works to interpret this relatively centrally dominated distribution in light of expectations from different FRB progenitor models, e.g.\ using the methods of \citet[][]{Bhandari+22}.

\section{Discussion}
\label{sec:discussion}

We have developed a combined framework for simultaneously fitting radio FRB data and optical follow-up images in \S~\ref{sec:xopt}, shown that it correctly estimates simulated truth parameters in \S~\ref{sec:sandbox}, applied it to a sample of CRAFT FRBs in \S~\ref{sec:craft} --- and found that it yields very little new information. However, the methodology presented here was explicitly developed to consider cases where there is significant host uncertainty, and the CRAFT ICS sample has the least in the entire FRB literature. We therefore now discuss cases where we expect this methodology to have a significant impact.

\subsection{Overcoming completeness cuts in \zdm}

Previous work in the \zdm\ code relied on analysing a `complete' section of the $z$--DM parameter space, represented by confident $P(U|x) \gtrsim 0.95$ host galaxies estimated by PATH. This could create a bias in the z--DM relation, since distant (high-$z$) FRBs would be less likely to have confidant hosts. While we have shown in \S~\ref{sec:bias_estimate} that such biases with current FRB samples are likely to be small, confirming the conclusions of \citet{2026ApJ...999..202S}, the actual bias in true FRB population parameters (as opposed to our simulated estimates of the truth) remains unknown.

The lower the DM of an FRB however, the lower the maximum FRB redshift, and hence the lower the chance of the host redshift distribution being biased. To ensure a complete sample therefore, in previous works using \zdm, a maximum FRB DM, DM$_{\rm EG}^{\rm max}$, was identified for each FRB survey, such that all FRBs with DMs up to and including DM$_{\rm EG}^{\rm max}$ would have confidant hosts (except where reasons unrelated to an FRB host's redshift, e.g.\ missing radio data or Galactic extinction, prevented optical follow-up). For the CRAFT ICS sample of \citet{Shannon_ICS}, we have set DM$_{\rm EG}^{\rm max}=1000$\,\pccc, due to the host of FRB\,20210912A, with DM$_{\rm EG} \approx 1150$, being unidentified \citep{Marnoch2023}. Consequently, the measured redshift of $1.016$ of FRB\,20220610A \citep{2024ApJ...963L..34G}, with DM$_{\rm EG} \approx 1380$\,\pccc\ \citep{Ryder2023}, has not been used in \zdm\ analyses.

This has proven to be a similarly minor effect for MeerKAT. \citet{PastorMarazuela2025} published the sample of 15 FRBs detected by MeerKAT, of which 13, when combined with FRB\,20230808F published by \citet{2025MNRAS.538.1800H}, represents a complete sample of all MeerKAT FRBs detected during 2023. Of these, nine have arcsecond localisations, yielding four confidant hosts with spectroscopic redshifts, and three with photometric redshifts. Of the two FRBs without confident hosts, one is located in the Galactic plane, so that ${\rm DM}_{\rm EG}^{\rm max} = 1395$\,\pccc\ is set by FRB\,20230827E, for which no host was found. However, this cut does not reduce the amount of available data, since neither FRB with higher \DMeg\ had an identifiable host. Four of the 13 FRBs have arcminute-scale localisations, and do not have confidant hosts.

Our first example where such completeness cuts severely impact the usable data volume is for DSA. \citet{2025PASA...42...17H} analysed the sample of 25 DSA FRBs presented by \citet{dsa_polarimitry_2024}, of which 11 had confidant hosts and measured redshifts \citep{Law2023}. FRB\,20220121B, with ${\rm DM}_{\rm EG} \approx 183$, did not have a measured redshift, setting DM$_{\rm EG}^{\rm max}=183$\,\pccc. This cut excluded eight of the redshifts from the analysis. The low value of DM$_{\rm EG}^{\rm max}$ was due to the relatively shallow Pan-STARRS1 catalogue \citep{ps1} being used for host identification.

An even more extreme example is that of CHIME FRBs, where the $\sim 10^\prime$ localisation accuracy of the majority of the 3641 unique sources \citep{CHIMECat2} precludes confidant host associations. The lack of host galaxy associations within the local volume for low-DM FRBs in the first CHIME catalog \citep{chimefrb_collaboration_first_2021,2025ApJ...989..130B} constrains the low end of the FRB energy distribution --- however, this non-observation, like other non-observations of FRB host candidates of sufficient nearness and/or brightness to be significant, could not be comprehensively included in a z--DM analysis. The sample of 81 FRBs localised to approximately $2^{"} \times 60^{"}$ by \citet{2025ApJS..280....6C} has 21 securely identified hosts. The completeness of this sample is limited by FRB 20230913A, with DM$_{\rm EG} \approx 150$\,\pccc, meaning that only 10 of the 21 FRBs with hosts lie within the ${\rm DM}_{\rm EG} < {\rm DM}_{\rm EG}^{\rm cut}=150$ region.

Our new methodology dispenses with the notion of completeness cuts, either in DM-space, or in $P(O|x)$, by fully modelling the bias against detecting more-distant FRB hosts. This will allow not just the many FRBs with confident FRB host associations, but which fail the $ {\rm DM}_{\rm EG} < {\rm DM}_{\rm EG}^{\rm cut}$ criteria, to be included in analyses --- it will also allow the information of a non-observation of a confident host, or the properties of ambiguous hosts, to be included in constraining FRB host and population models.

We have not included such a re-analysis of the above samples in this current work, purely because we wish to focus on the methodology. We instead leave this to future works, where we will aim to include MeerTRAP FRBs with arcminute localisations, DSA FRBs with and without confidant hosts, and a very large number of FRBs from CHIME, potentially ranging from VLBI localisations using CHIME outriggers, to those without even baseband data.

\subsection{Inclusion of repeating FRBs}

The question of whether or not all FRBs repeat has been the subject of much debate --- a good discussion covering recent results is given in \citet{2026arXiv260508410C}. For now, we highlight the suggestion that repeating FRBs favour dwarf, low-metallicity hosts, while apparently once-off FRBs more closely track star-formation. The first evidence of this came when once-off FRB~20180924A was localised to a large spiral galaxy \citep{bannister_single_2019}, in contrast to the dwarf host of repeating FRB~20121102A \citep{2017Natur.541...58C,121102Host}.  Statistical tests of host properties of repeaters and once-offs, e.g.\ those performed \citet{Bhandari+22}, have been hampered however by the localisation bias. It is much easier to localise, and hence find the host galaxies for, repeating FRBs, both because multiple detections further constrain the localisation region \citep[as was used to identify NGC 3252 as the host of FRB\,20181030A by][]{2021ApJ...919L..24B}, and because they can be followed up with more-sensitive, higher angular resolution telescopes than the discovery instrument \citep[for example, the localisation of the first repeater, FRB~20121102A][]{2017Natur.541...58C,121102Host}. Thus, dwarf hosts can be pinpointed much more securely for repeating than once-off FRBs, as most obviously demonstrated by the confirmation of the lowest-mass FRB host galaxy yet, being the $\log_{10}(M_*/M_\odot) = 7.88$ dwarf host for repeating FRB\,20190417A by \citet{2024ApJ...976..199I,2026ApJ...996L..16M}.

Repeating FRBs are expected to have different redshift distributions than non-repeating FRBs, since they must necessarily be closer for multiple bursts to be detectable \citep{Gardinier2021_frbpoppy_repeaters}. FRBs observed to repeat more frequently will also be more likely to have a known host, because follow-up observations will be more likely to detect subsequent bursts, as well as having close hosts. The simplest demonstration of this is FRB~20180916B \citep{2020Natur.577..190M}, which remains the CHIME FRB with the greatest number of repeats seen \citep{2026arXiv260508410C}, and has a host galaxy at only $z=0.0337$.

A simple implementation of repeating FRBs, which assumes Poissonian arrival times, and time/frequency structures which are independent of repetition strength, has been implemented in \zdm. It has been used to show that the declination and rate distribution of repeating FRBs, and the ratio of once-off to multiple bursts, in CHIME Catalogues 1 \citep{JamesRepeating2023} and 2 \citep{2026arXiv260508410C} are consistent with all FRBs being due to intrinsically repeating sources. However, to date, no repeating FRB redshift has been included in \zdm.  Until now, doing so would require identifying a completeness region in \DMeg\ and $N_{\rm rep}$ space, where all repeating FRBs with at least $N_{\rm rep} > N_{\rm rep}^{\rm cut}$ detected repeats, and $\DMeg < {\rm DM}_{\rm EG}^{\rm cut}$, have an identified host. However, when applied to the 16 repeaters identified from CHIME Catalogue 1, some FRBs with $N_{\rm rep}=3$ do not have confident hosts, so that only FRB~20180916B ($N_{\rm rep}=19$) and FRB~20180814A ($N_{\rm rep}=11$) pass the $N_{\rm rep} > N_{\rm rep}^{\rm cut}$ criteria.

The scarcity of repeating FRBs passing this completeness cut means that the redshift distribution of repeating FRBs is poorly constrained, since it is possible to fit the sample with a lower density of rapidly repeating objects, which are then detectable from a greater distance, and a large number of more rarely repeating progenitors, which are more likely to be located close enough to be detected \citep{JamesRepeating2023}.

Within our framework, the number of observed repeats, $N_{\rm rep}$, becomes a measureable radio property, i.e.\ $N_{\rm rep} \in \xrad$. Using repeating FRBs therefore requires modelling $P(N_{\rm rep})$ and $P(z|N_{\rm rep})$ --- which the \zdm\ code is already capable of doing. Thus, our framework should allow the known redshifts of repeating FRBs to be used to constrain their distribution. Such an analysis could also allow tests of whether or not the host galaxy distribution of repeating FRBs is significantly different from that of apparently once-off FRBs. Since our code accounts for this bias, it should be able to reliably test (for instance) whether or not the true host magnitude distributions are consistent between repeaters and apparent once-offs.

There is one complication to this optimism however. In order to localise most repeaters, follow-up observations need to detect at least one burst.
Care will therefore have to be taken to account for any such biases arising from the follow-up procedure. We also leave such considerations to a future work.

\subsection{Biases we cannot correct for}

There is one remaining biases in the FRB literature that our new methodology cannot overcome --- and that is reporting bias. There is a tendency, especially within astronomical literature, to only publish positive results, and not report observations where nothing is seen. In the case of FRB literature, this can result in preferential publication of FRBs with unusual host galaxy properties, with an example being \citet{Ryder2023}, where the high ($z>1$) redshift of the host led to publication before several less interesting FRBs from the same survey were made public. Another example is the 40 DSA FRBs presented by \citet{DSA_Sharma_sfr} and \citet{2025NatAs...9.1226C}. In those works, only FRBs with confident $(P(U|x)>0.95)$ host associations were published, leading to expected biases towards low-redshift, bright host galaxies. However, since FRBs without confident hosts were not published, completeness conditions could not even be estimated.

In terms of our framework, these and similar works are choosing to include $P(\xopt|\xrad)$ based on the observed properties of $\xopt$ --- which opens one up to bias. As noted previously, one is free to choose any set of selection cuts based on $\xrad$, and include $P(\xopt|\xrad)$ based on these, provided that the distribution $P(\xopt|\xrad)$ accounts for those selection effects. A good example would be removing FRBs detected at low Galactic latitudes, which should not correlate at all with FRB host properties, but does significantly increase uncertainty in \DMeg, and reduces the fidelity of follow-up images. Another is the aforementioned calculation of $P(\xopt|N_{\rm rep} \in \xrad)$, where the observation of $N_{\rm rep}$ repeat bursts, required to include optical data, is explicitly accounted for.

\subsection{Deficiencies in the current work}

The key result of this is \S\,\ref{sec:xopt}, which merely presents the framework for simultaneously accounting for both optical and radio biases in FRB host galaxy analysis. However, our specific implementation of this framework is imperfect, and we here highlight several areas which we consider should be targeted for improvements.

Firstly, there are the deficiencies in the three host galaxy models presented in \citet{zdm_path_hosts}. As discussed in that work, the only host model based on theoretical expectations \citep[the Loudas25 model, based on scaling FRBs with host star-formation and/or stellar mass as per ][]{Loudas25} only fits the expected FRB host galaxy distribution when taken to an unphysical limit ($f_{\rm sfr} \gg 1)$. The other, better-fitting models are empirical, and based on FRB data mostly in the $z \lesssim 0.5$ range, and thus may not extrapolate well to higher redshifts. We have also specified these models in terms of R-band magnitude, $m_R$, only --- ideally, they should be adaptable to any commonly used optical band. We strongly encourage the community to continue to improve on FRB host models.

Secondly, our current model for $P_F(z|m)$ (i.e., the redshift distribution of field galaxies) is only approximate, and is poorly estimated above $z=2$, which affects $P_F(z|m)$ when $m \gtrsim 20$. This could be improved with the use of deeper spectroscopic catalogues, which we aim to implement in future works.

Thirdly, in our methodology, host models are treated independently of FRB population properties. However, in reality, they will not be. For instance, a host model with FRBs tracing star-formation ($f_{\rm sfr}=1$) should also imply that the FRB population evolves with the star-formation rate ($\nsfr=1$), while a fit for FRBs originating from the centres of hosts (i.e., low values of $\theta_0$) should also yield higher expected values of \DMhost. Currently, it is probably best to keep these parameters independent, which allows fits that produce inconsistent values to be discarded, and greater emphasis to be placed on those results that tell a consistent story. However, as FRB analysis has grown more sophisticated, so has the number of correlated parameters being fit, meaning that the precision on measurements of key properties, such as $H_0$, has not greatly improved with time. It may therefore be necessary to reduce the number of degrees of freedom by linking such analyses in the future.

Fourthly, there have been several works suggesting correlations between FRB properties and hosts --- in particular, higher scattering is expected to correlate with higher \DMhost\ \citep{CordesTauRedshift2022}, though such a relation is not seen empirically \citep{CRAFT_HTR}, possibly due to observational biases and fluctuations in \DMcosmic\ \citep{2026ApJ...998....1M}. Correlations have also been observed between a host's inclination and an FRB's DM \marnoch\ and rotation measure, RM \citep{2025PASA...42..157G}. Including such correlations into our analysis would be possible, but would require adding RM to \xrad\ ($\tau$ is already included, albeit not used in this work) and inclination to $\xopt$, with a resulting increase in the complexity of the analysis.

Fifthly and lastly, our method of generating synthetic images only includes FRB hosts in the magnitude range $14 \le m_r \le 22$, which limits our ability to simulate real FRB experiments. Furthermore, it only allows us to know the redshift of the true host. This should not be a problem if the statistical derivation of \S\,\ref{sec:xopt} is correct, but it does prevent us fully testing the effect of $P(z|m)$ terms in Eq.~\ref{eq:full_prob}. We suggest performing  a dedicated analysis, similar to the one performed by \Andersen\ for the \pth-only case, but with full redshift information.

\section{Conclusion}

We have developed a methodology that fully accounts for uncertainty in the host galaxies of FRBs when estimating both host galaxy and FRB population parameters. We have extended previous work that provides well-motivated priors on FRB host magnitudes, using the FRB z--DM (Macquart) relation and host galaxy models, to produce a joint probability distribution of both radio and optical observables. We have so-far included only FRB host galaxy magnitude and redshift in the optical observables, but our methodology could be extended to include other parameters, such as angular size, or morphology.

We have implemented our methodology by combining the \zdm\ and \pth\ codes. Tests using a synthetic sample of FRBs shows that both FRB host galaxy, and FRB population parameters, can be estimated accurately, with biases only occurring because of small image sizes.
We have found correlations between FRB host galaxy properties and FRB population parameters to be less than 10\%, and little bias in FRB population parameters when using confidant hosts only in a traditional \pth\ analysis, for a synthetic sample of well-localised FRBs. Nonetheless, such a bias may become important for precision cosmological measurements using more than 1,000 FRBs --- and our methodology negates this, except for the inevitable bias due to the influence of the priors used in the analysis.

We find that the main advantages of our method are to be able to directly fit for FRB host and population parameters, while fully and consistently accounting for biases; to
be able to use poorly localised FRBs (i.e., with $30"$ localisation uncertainty) to study FRB host galaxies, where traditional methods, using confidant FRB hosts, fail; to improve confidence in uncertain FRB host associations, as demonstrated by confirming the previously identified host of FRB\,20190611B; and (least significantly) to improve statistical confidence in parameter estimates, due to the ability to include more data.

Applied to the CRAFT sample of FRBs, we find both FRB host and population parameters consistent with previous results, due to the high localisation accuracy of radio measurements, and deep optical follow-up. This also allows us to fit the FRB offset distribution, finding that FRBs are strongly biased towards the centres of their hosts, consistent with an exponential surface density with scale factor $0.41^{+0.13}_{-0.09}$ times the half-light radius.

Our methodology was developed however for cases where deep optical follow-up is not possible, and/or the FRB localisation region is insufficient to clearly identify the host galaxy. We hope to apply our model to such data in the future. We have also identified several deficiencies in the current work, and highlight the need to test our model with a more detailed synthetic host galaxy sample, and to improve our models of FRB hosts. Future extensions to the work might include adding priors according to other host properties, e.g.\ incorporating correlations between DM and host galaxy inclination (\marnoch); we have been made aware, for instance, that Lee et al.\ (in prep) make use of spectroscopic foreground data \citep[e.g.\ as per ][]{2022ApJ...928....9L,FLIMFLAMdr1} as conditional information for simultaneous inference of PATH posteriors, host DMs, and IGM/halo gas fractions.

\begin{acknowledgement}

This scientific work uses data obtained from Inyarrimanha Ilgari Bundara, the CSIRO Murchison Radio-astronomy Observatory. We acknowledge the Wajarri Yamaji People as the Traditional Owners and native title holders of the Observatory site. CSIRO’s ASKAP radio telescope is part of the Australia Telescope National Facility (https://ror.org/05qajvd42). Operation of ASKAP is funded by the Australian Government with support from the National Collaborative Research Infrastructure Strategy. ASKAP uses the resources of the Pawsey Supercomputing Research Centre. Establishment of ASKAP, Inyarrimanha Ilgari Bundara, the CSIRO Murchison Radio-astronomy Observatory and the Pawsey Supercomputing Research Centre are initiatives of the Australian Government, with support from the Government of Western Australia and the Science and Industry Endowment Fund.

This work was performed on the OzSTAR national facility at Swinburne University of Technology. The OzSTAR program receives funding in part from the Astronomy National Collaborative Research Infrastructure Strategy (NCRIS) allocation provided by the Australian Government, and from the Victorian Higher Education State Investment Fund (VHESIF) provided by the Victorian Government.

Based on observations collected at the European Organisation for Astronomical Research in the Southern Hemisphere under ESO programme 1108.A-0027(A).
\end{acknowledgement}

\paragraph{Funding Statement}

The authors declare no external sources of funding.

\paragraph{Competing Interests}

The authors declare no competing interests.

\paragraph{Data Availability Statement}

The data and results from this work can be obtained from the \zdm\ GitHub codebase \citep{zdm}, at \url{https://github.com/FRBs/zdm}.

\printendnotes

\printbibliography

\appendix

\section{Estimating \emin}
\label{sec:emin}

Within \zdm, the two implemented luminosity functions are a power-law, and a Schechter function, with the Schechter function having a smooth exponential cut-off at high energies. Since \citet{2025PASA...42...17H}, works using \zdm\ have used the Schechter function, simply because the sharp cut-off of the power-law at \emax\ was seen as unphysical --- even though there is no direct experimental evidence for any kind of cut-off, exponential or otherwise, either from non-parametric studies of the energy distribution \citep{2025PASA...42....3A}, or from detailed studies of repeating FRBs \citep{niu_repeating_2022,2022RAA....22l4002Z,2026MNRAS.545f1937O}.
However, both functions have a sudden, sharp cut-off at \emin, below which the probability of observing an FRB becomes zero. In the first \zdm\ paper, \emin\ was set to $\lemin = 10^{30}$ --- well below the energy probed by the data --- since it was shown that there was no evidence for a low-energy cut-off in FRB emission; a 90\% confidence upper limit on $\lemin = 39.0$ was set \citep{james_zdm_2022}. Due to the deficit in low-DM FRBs observed by FAST however \citep{Niu2021}, \citet{2025PASA...42...17H} re-introduced \emin\ as a minimisation parameter, with the optimisation finding $\lemin=39.47_{-1.28}^{+0.54}$ at 68\% confidence. This was interpreted not as a literal minimum energy, but instead as a potential flattening of the FRB luminosity function at low energies, since repeating FRBs had been observed to emit at significantly lower energies \citep[e.g.][]{2021Natur.598..267L}.

\begin{figure}
    \centering
    \includegraphics[width=\linewidth]{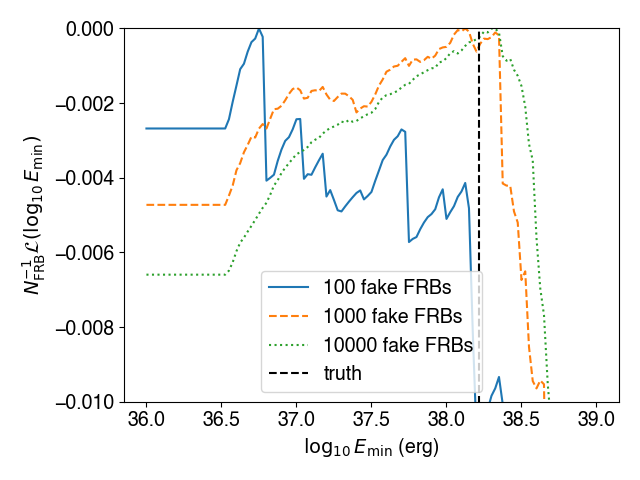}
    \caption{Likelihood of samples of fake FRBs as a function of \lemin, normalised by the number of FRBs used. The simulated truth value is also given.}
    \label{fig:emin}
\end{figure}

Here, we use our Monte Carlo FRB sample to determine the ability of \zdm\ to correctly identify such an \emin. We generate FRBs as described in \S~\ref{sec:sandbox}, albeit using the value of $\lemin=38.22$ found by \citet{HoffmannHalo}. We then estimate $\mathcal{L}(\emin)$ as a function of \emin\ only using Eq.~\ref{eq:full_prob}, with all other parameters set to their true values. The result is shown in Figure~\ref{fig:emin}.

When using 10,000 FRBs, we find that the resulting likelihood distribution is well-behaved, and peaks at the true value. However, for small samples of FRBs, neither is true -- for 1,000 FRBs, the likelihood fluctuates, while for 100, it experiences large, sudden jumps. In reality of course, the apparently smooth behaviour of the 10,000 FRB distribution is due to averaging over a very large number of such jumps.

The cause of this behaviour is the manner in which the function $F(B,\weff,s)$ from Eq.~\ref{eq:F} is calculated. In reality, all three variables --- $\weff$, $B$, and $s$ --- have a smooth probability distribution. However, within \zdm, the distributions $p(B)$ and $p(\weff)$ are parameterised by histograms, with bin values $b_1, b_2, \ldots, b_{N_b}$ and $w_1, w_2, \ldots, w_{N_w}$, where typically $N_b \sim N_w \sim 10$. Any given FRB with measured $s$ will therefore have a finite number possible fluences $F(B,\weff,s)$, corresponding to all possible combinations of width and beam bin. If $B$ and $\weff$ are also measured for that FRB, then its likelihood will be calculated via a weighted sum over the two $b$ and two $w$ bins bracketing its measured values. As \emin\ increases, it will generally increase the likelihood of all FRBs, except when \emin\ exceeds the implied energy corresponding to some combination of $B$ and $w$ for a particular FRB. At that point, the likelihood will suddenly drop, since it will lose the probability mass associated with those bins in $B$ and $w$. And as Figure~\ref{fig:emin} shows, a very large number of FRBs is required for this behaviour to average out to produce a well-behaved likelihood.

The ideal solution to this behaviour is to introduce a luminosity function without a sharp cut-off at low-energies --- either an exponential turn-on, or a broken power-law, seem reasonable possibilities. However, extending \zdm\ analysis to further luminosity variables is beyond the scope of the current work. Hence, for now, we recommend setting $\lemin=30$ (or some other very low number) for purposes of modelling the FRB population.

\end{document}